 \documentclass[preprint,aps]{revtex4} 

\usepackage{graphicx}
\newcommand {\bea}{\begin{eqnarray}}
\newcommand {\eea}{\end{eqnarray}}
\newcommand {\be}{\begin{equation}}
\newcommand {\ee}{\end{equation}}
\newcommand {\pslash}{p\!\!\!/}

\newcommand {\Dslash}{D\!\!\!/}
                      
\begin{document}


\title{Instantons in QCD with Many Colors}

\author{T.~Sch\"afer$^{1,2,3}$}

\affiliation{
$^1$Department of Physics, Duke University, Durham, NC 27708\\
$^2$Department of Physics, SUNY Stony Brook, Stony Brook, NY 11794\\ 
$^3$Riken-BNL Research Center, Brookhaven National Laboratory, 
Upton, NY 11973}

\begin{abstract}
 We study instantons in QCD with many colors. We first discuss
a number of qualitative arguments concerning the large $N_c$
scaling behavior of a random instanton ensemble. We show that
most hadronic observables are compatible with standard large 
$N_c$ counting rules provided the average instanton size is 
$O(1)$ and the instanton density is $O(N_c)$ in the large $N_c$ 
limit. This is not the case for the topological susceptibility 
and the mass of the $\eta'$. For these observables consistency
with conventional large $N_c$ counting requires that fluctuations
in the instanton liquid are suppressed compared to Poissonian 
fluctuations. Using mean field estimates and numerical
simulations we show that the required scaling behavior of the 
instanton density is natural in models in which the instanton 
density is regularized in terms of a classical repulsive core. 
We also show that in these models fluctuations of the topological 
charge are suppressed and that $m^2_{\eta'}=O(1/N_c)$. We
conclude that the instanton liquid model is not necessarily in 
conflict with the $1/N_c$ expansion. 

  
\end{abstract}

\maketitle
\newpage

\section{Introduction}
\label{sec_intro}

 Quantum Chromodynamics (QCD), the theory of the strong interactions, 
is an essentially parameter free theory. The QCD lagrangian contains
a coupling constant, $g$, but because of the phenomenon of dimensional
transmutation this parameter is replaced by a dimensionful scale 
parameter, $\Lambda_{QCD}$. As a consequence, there is no obvious
expansion parameter in QCD. 't Hooft proposed to use the number 
of colors, $N_c$, as a parameter \cite{'tHooft:1973jz}. He suggested
that QCD simplifies in the limit $N_c\to\infty$ and that $1/N_c$
can be treated as an expansion parameter. 

 We do not know how to solve QCD in the large $N_c$ limit. However,
by analyzing the $N_c$ dependence of classes of Feynman diagrams we
can make certain qualitative statements about the structure of 
large $N_c$ QCD. We expect, for example, that the masses of
mesons and glueballs are $O(1)$ whereas the masses of baryons 
are $O(N_c)$. Also, meson decay constants are $O(N_c^{1/2})$
and meson-meson scattering amplitudes are $O(1/N_c)$. As a
result, large $N_c$ QCD is a theory of weakly interacting
mesons and glueballs. 

 The fate of the $U(1)_A$ anomaly in the large $N_c$ limit
is an interesting question. The problem is complicated by 
the fact that if the $\theta$ term,
\be
 {\cal L} = \frac{ig^2\theta}{32\pi^2}G_{\mu\nu}^a\tilde{G}_{\mu\nu}^a,
\ee
is added to the QCD lagrangian then in perturbation theory 
there is no dependence on the parameter $\theta$. Witten suggested 
that non-perturbative effects generate $\theta$-dependence in 
the pure gauge theory and that the topological susceptibility, 
\be
\label{chi_top}
 \chi_{top} = \left.\frac{d^2E}{d\theta^2}\right|_{\theta=0},
\ee
is $O(1)$ in the large $N_c$ limit \cite{Witten:1978bc,Witten:1998uk}.
Standard large $N_c$ counting suggests that the contribution of 
fermions to $\chi_{top}$ is subleading in $1/N_c$. We know, however,
that this is not correct, because there is no $\theta$-dependence 
in QCD with massless fermions. Witten argued that this apparent 
contradiction can be resolved
if the mass of the $\eta'$ meson scales as $N_c^{-1/2}$ in the
large $N_c$ limit. Witten \cite{Witten:1979vv} and Veneziano 
\cite{Veneziano:1979ec} derived a relation between the mass 
of the $\eta'$ and the topological susceptibility in pure gauge 
theory,
\be 
\label{WV}
\frac{f_\pi^2}{2N_f} m_{\eta'}^2 = \chi_{top}.
\ee
Using $\chi_{top}=O(1)$ and $f_\pi^2=O(N_c)$ we observe
that indeed $m_{\eta'}^2=O(1/N_c)$. This result implies
that the $U(1)_A$ anomaly is effectively restored in the
large $N_c$ limit. 

 The $\theta$-dependence of vacuum energy is related to
topological properties of QCD. In the semi-classical 
approximation these features can be described in terms
of instantons. Instantons are localized field configurations
that carry topological charge \cite{Belavin:fg}
\be 
 Q_{top} = \frac{g^2}{32\pi^2}\int d^4 x\,
  G_{\mu\nu}^a\tilde{G}_{\mu\nu}^a = \pm 1.
\ee
In 1978 Witten observed that classical effects, such as 
instantons, scale as $\exp(-1/g^2)\sim \exp(-N_c)$ which 
seems to contradict the assumption $\chi_{top}=O(1)$ 
\cite{Witten:1978bc}. On the other hand, it was also found 
that the assumption that topological fluctuations are 
semi-classical leads to a very successful picture 
of the QCD vacuum, termed the instanton liquid model
\cite{Callan:1977gz,Shuryak:1981ff,Diakonov:1983hh,Schafer:1996wv}. 
The instanton liquid model postulates that for $N_c=3$ 
the density of instantons is approximately 
$(N/V)\simeq 1\, {\rm fm}^{-4}$ while the average size is 
$\rho\simeq 1/3\, {\rm fm}$. These numbers reproduce the 
topological susceptibility in the pure gauge theory 
$\chi_{top}\simeq (200\, {\rm MeV})^4$ and the chiral 
condensate $\langle \bar{\psi}\psi\rangle \simeq -(230\,
{\rm MeV})^3$. More detailed calculations show that the 
instanton liquid model successfully describes an impressive 
amount of data on hadronic correlation functions 
\cite{Shuryak:1993,Schafer:1996wv,Schafer:2000rv}.
 
  Topological properties of the QCD vacuum have also been 
studied in lattice QCD. It was found that the topological 
susceptibility in pure gauge QCD is $\chi_{top}\simeq (200\, 
{\rm MeV})^4$ \cite{Teper:1999wp}, as predicted by the Witten-Veneziano 
relation equ.~(\ref{WV}). However, it was also observed that the 
topological susceptibility is very stable under cooling, and 
appears to be dominated by semi-classical configurations. 
Lattice simulations also seem to confirm the values of the 
key parameters of the instanton liquid \cite{Chu:1994}, $(N/V)
\simeq 1\, {\rm fm}^{-4}$ and  $\rho\simeq 1/3\, {\rm fm}$. 
Recent work has focused on the relation of instantons 
and low-lying eigenstates of the Dirac operator. The instanton 
model predicts that the lowest eigenstates of the Dirac operator, 
which dominate chiral symmetry breaking, are linear combinations 
of localized, approximately chiral states associated with the 
fermionic zero modes of individual instantons and anti-instantons
\cite{Diakonov:1985}. This picture has been confirmed by 
lattice calculations \cite{DeGrand:2000gq,Horvath:2001ir}.

 In light of these results the question arises whether the 
success of the instanton model can be reconciled with the large 
$N_c$ expansion. In order to address this problem we have 
investigated the predictions of the instanton model for QCD 
with many colors. This paper is organized as follows. In 
Sect.~\ref{sec_ril} we present qualitative arguments 
based on a random instanton liquid. In Sects.~\ref{sec_mfa} 
and \ref{sec_csb} we discuss analytic results obtained
in the mean field approximation. In Sects.~\ref{sec_ilm}
and \ref{sec_cor} we discuss the results of numerical
simulations of the interacting instanton liquid.

\section{Random Instanton Liquid}
\label{sec_ril}

  The simplest picture of the instanton liquid in QCD is 
the random instanton liquid model proposed by Shuryak 
\cite{Shuryak:1981ff}. This model is based on the assumption 
that the instanton ensemble is characterized by two key parameters, 
the instanton density $(N/V)$ and the typical instanton size $\rho$. 
Except for the size, the collective coordinates of the instantons 
are distributed randomly. The model refers to an instanton liquid 
rather than an instanton gas because interactions between instanton 
cannot be ignored completely. This is particularly clear if massless
fermions are present. If interactions are ignored then every instanton 
contributes an exact zero mode to the spectrum of the Dirac operator 
and the total density of instantons is exactly zero. Interactions 
lead to mixing between the zero modes and qualitatively change the 
spectrum of the Dirac operator. As a consequence, chiral symmetry
is spontaneously broken and the density of instantons is non-zero.

  In order to understand the large $N_c$ behavior of the 
instanton liquid we have to understand the scaling behavior
of the two parameters introduced above. In this paper we 
shall argue that 
\be
\label{nc_as}
 \frac{N}{V}=O(N_c), \hspace{1cm} \rho = O(1).
\ee
In section \ref{sec_mfa} we will show that equ.~(\ref{nc_as})
follows from the partition function of the instanton 
liquid if the instanton size distribution is regularized 
by a classical repulsive core. Before we discuss this
result we would like to provide some plausibility
arguments and explore simple consequences of the scaling 
relations. 

 We first consider the average instanton size $\rho$. At 
one-loop order the instanton action is given by $S_0=(8\pi^2)/
g^2=-b\log(\rho\Lambda)$ where $b=(11N_c)/3$ is the first 
coefficient of the beta function in pure gauge QCD. In the 
't Hooft limit $N_c\to\infty$ with $g^2N_c=const$ we expect 
$S_0=O(N_c)$ and $\rho=O(1)$. We have to keep in mind, however, 
that instantons in QCD come in all sizes. In the semi-classical
approximation we cannot study instantons with size $\rho\sim 
\Lambda^{-1}$ and action $S_0\sim 1$. There is evidence, 
however, that chiral symmetry breaking in QCD with $N_c=3$ colors 
is dominated by small instantons. A priori it is not clear
whether this remains true in the large $N_c$ limit. Our 
strategy in this work is to assume that the semi-classical
approximation remains valid and to show that this assumption 
leads to a consistent picture. 

 In the random instanton liquid model the instanton density is related 
to the non-perturbative gluon condensate 
\be 
\label{glue}
 \frac{N}{V} = \frac{1}{32\pi^2}
   \langle g^2 G^a_{\mu\nu} G^a_{\mu\nu} \rangle .
\ee
Standard $N_c$ counting suggests that $\langle g^2G^2\rangle =
O(N_c)$ and we are lead to the conclusion that $(N/V)=O(N_c)$. 
This is also consistent with the expected scaling of the vacuum 
energy. Using equ.~(\ref{glue}) and the trace anomaly relation 
\be 
\label{trace_anom}
\langle T_{\mu\mu} \rangle = -\frac{b}{32\pi^2}
  \langle g^2 G^a_{\mu\nu} G^a_{\mu\nu} \rangle ,
\ee
the vacuum energy density is given by
\be
\label{eps}
\epsilon = -\frac{b}{4} \left(\frac{N}{V}\right).
\ee 
Using $(N/V)=O(N_c)$ we find that the vacuum energy scales as 
$\epsilon = O(N_c^2)$ which agrees with our expectations for a 
system with $N_c^2$ gluonic degrees of freedom. We should note 
that equ.~(\ref{eps}) requires interactions between instantons.
The partition function of a completely non-interacting gas
of instantons gives $\epsilon\sim (N/V)\sim N_c$. 

  We also note that $(N/V)=O(N_c)$ implies that the effective 
``packing fraction'' of instantons remains constant in the 
large $N_c$ limit. Instantons with topological charge $Q_{top}
=\pm 1$ in $SU(N_c)$ QCD are embeddings of $SU(2)$ instantons. 
Since the number of mutually commuting $SU(2)$ subgroups of 
$SU(N_c)$ scales as $N_c$ we can have $O(N_c)$ instantons
which overlap in space but are in fact weakly interacting.
Witten argued that the only alternatives are $S_0=O(N_c)$
and $(N/V)=O(e^{-N_c})$ or $S_0=O(1)$ and $(N/V)=O(e^{N_c})$. 
However, as we shall see in the next section, it is possible 
for the instanton density to remain finite even if $S_0=
O(N_c)$ because the large entropy of instantons in $SU(N_c)$
can overcome the exponential suppression due to the action.

 If instantons are distributed randomly then fluctuations
in the number of instantons and anti-instantons are expected 
to be Poissonian. This leads to the predictions
\bea
\label{n2}
 \langle N^2 \rangle - \langle N\rangle^2 
  = \langle N \rangle , \\
\label{q2}
 \langle Q^2 \rangle  =  \langle N \rangle ,
\eea
where $N=N_I+N_A$ is the total number of instantons and 
$Q=N_I-N_A$ is the topological charge. Equ.~(\ref{q2}) 
implies that 
\be
 \chi_{top} = \frac{\langle Q^2\rangle}{V} = 
   \left(\frac{N}{V}\right) .
\ee
Using $(N/V)=O(N_c)$ we observe that $\chi_{top}=O(N_c)$ which is 
in contradiction to Witten's assumption $\chi_{top}=O(1)$. However,
as we shall see in the next section, interactions between instantons
cannot be ignored in the large $N_c$ limit and the fluctuations 
are suppressed compared to equs.~(\ref{n2},\ref{q2}). 

 Finally, we would like to study chiral symmetry breaking 
in a random instanton liquid. For definiteness, we will consider
the case $N_f=2$ but the conclusions are of course independent 
of the number of flavors. Instantons induce an effective 
$2N_f$-fermion lagrangian. After averaging over the color
orientation of the instanton the effective lagrangian is given by
\cite{'tHooft:up,Shifman:uw}
\bea
\label{l_nf2}
{\cal L} &=& \int n(\rho)d\rho\, 
   \frac{2(2\pi\rho)^4\rho^2}{4(N_c^2-1)}
 \epsilon_{f_1f_2}\epsilon_{g_1g_2}
 \left( \frac{2N_c-1}{2N_c}
  (\bar\psi_{L,f_1} \psi_{R,g_1})
  (\bar\psi_{L,f_2} \psi_{R,g_2}) \right. \\
& & \hspace{1cm}\mbox{}\left. - \frac{1}{8N_c}
  (\bar\psi_{L,f_1} \sigma_{\mu\nu} \psi_{R,g_1})
  (\bar\psi_{L,f_2} \sigma_{\mu\nu} \psi_{R,g_2})
  + (L \leftrightarrow R ) \right) \nonumber .
\eea
We observe that the explicit $N_c$ dependence is given by 
$1/N_c^2$. This is again related to the fact that instantons
are $SU(2)$ objects. Quarks can only interact via instanton 
zero modes if they overlap with the color wave function of
the instanton. As a result, the probability that two quarks 
with arbitrary color propagating in the background field of
an instanton interact is $O(1/N_c^2)$. 

 Chiral symmetry breaking can be studied in the mean field
approximation. We will address this problem in much more detail
in the following section but we can give a simple qualitative 
argument here. In the mean field approximation we can derive 
a gap equation for the spontaneously generated constituent 
quark mass. The gap equation is of the form
\be 
M = GN_c \int \frac{d^4k}{(2\pi)^4}\frac{M}{M^2+k^2},
\ee
where $M$ is the constituent mass and $G$ is the effective 
coupling constant in equ.~(\ref{l_nf2}). The factor $N_c$
comes from doing the trace over the quark propagator. The 
coupling constant $G$ scales as $1/N_c$ because the density of 
instantons is $O(N_c)$ and the effective lagrangian contains an 
explicit factor $1/N_c^2$. We conclude that the coefficient in 
the gap equation is $O(1)$ and that the dynamically generated
quark mass is $O(1)$ also. This also implies that the quark
condensate, which involves an extra sum over color, is 
$O(N_c)$.

\section{The Mean Field Approximation}
\label{sec_mfa}


  In this section we shall make the arguments presented in the 
previous section more quantitative. Instead of considering 
the average instanton size and instanton size density to be 
arbitrary parameters, we shall determine these quantities
from the partition function of the instanton ensemble. This
means, in particular, that both $\rho$ and $(N/V)$ are 
expressed in terms of the fundamental scale parameter
of QCD. For this purpose it is essential to take the 
interaction between instantons into account. If instantons
are semi-classical, $S_{inst}\gg 1$, and if the interaction 
between instantons is weak, $S_{int}\ll S_{inst}$, this 
can be accomplished using mean field methods
\cite{Ilgenfritz:1980vj,Diakonov:1983hh,Munster:2000uu}. In this
section, we shall follow the variational method of 
Diakonov and Petrov \cite{Diakonov:1983hh}.

 We consider the partition function for a system of instantons 
in pure gauge theory
\be
\label{Z}
 Z = \frac{1}{N_I!N_A!}\prod_I^{N_I+N_A}\int [d\Omega_I\, n(\rho_I)]
 \, \exp(-S_{int}).
\ee
Here, $\Omega_I=(z_I,\rho_I,U_I)$ are the collective coordinates of the 
instanton $I$ and  $n(\rho)$ is the semi-classical instanton distribution
function \cite{'tHooft:fv}
\bea
\label{n(rho)}
  n(\rho) &=& C_{N_c} \ \left(\frac{8\pi^2}{g^2}\right)^{2N_c} 
 \rho^{-5}\exp\left[-\frac{8\pi^2}{g(\rho)^2}\right],\\
 && C_{N_c} \;=\; \frac{0.466\exp(-1.679N_c)}
    {(N_c-1)!(N_c-2)!}\, ,\\
 && \frac{8\pi^2}{g^2(\rho)} \;=\; 
    -b\log(\rho\Lambda), \hspace{1cm} 
    b = \frac{11}{3}N_c \, . 
\eea
We have denoted the classical instanton interaction by $S_{int}$. 
If the instanton ensemble is sufficiently dilute we can approximate
the instanton interaction as a sum of two-body terms, $S_{int}=
\sum_{IJ} S_{IJ}$. For a well separated instanton-anti-instanton
pair the interaction has the dipole structure \cite{Callan:1977gz}
\be
\label{int_dip}
S_{int} = -\frac{32\pi^2}{g^2} \frac{\rho_I^2\rho_A^2}{R^4_{IA}}
 |u|^2 \left( 1- 4\cos^2\theta \right).
\ee
Here $\rho_{I,A}$ are instanton radii and $R_{IA}$ is the 
instanton-anti-instanton separation. The relative color orientation
is characterized by a complex four-vector $u_\mu=\frac{1}{2i}{\rm tr}
(U_{IA}\tau^+_\mu)$, where $U_{IA}=U_I U_A^\dagger$ depends on 
the rigid gauge transformations that describe the color orientation
of the individual instanton and anti-instanton and $\tau^+_\mu=
(\vec\tau,-i)$. We have also defined the relative color angle 
$\cos^2\theta = |u\cdot\hat R|^2/|u|^2$. The dipole interaction
is valid if $R_{IA}^2\gg \rho_I\rho_A$. We will specify the 
interaction at shorter distances below. 

   Diakonov and Petrov suggested to analyze the partition function
equ.~(\ref{Z}) using a variational single-instanton distribution 
$\mu(\rho)$ \cite{Diakonov:1983hh}. The corresponding partition
function is 
\bea
\label{Z_var}
 Z_1 = \frac{1}{N_I!N_A!}\prod_i^{N_I+N_A}\int d\Omega_I\, 
 \mu(\rho_I)= \frac{1}{N_I!N_A!}(V\mu_0)^{N_I+N_A}
\eea
where $\mu_0 = \int d\rho\,\mu(\rho)$. The exact partition 
function is 
\be
\label{Z-Z1}
 Z = Z_1 \langle \exp(-(S-S_1)) \rangle, 
\ee
where $S$ is the full action, $S_1=\log(\mu(\rho))$ is the 
variational estimate and the average $\langle .\rangle$ is
computed using the variational distribution function. The 
partition function satisfies the bound 
\be
\label{bound}
 Z \geq Z_1 \exp(-\langle S-S_1\rangle ),
\ee
which follows from convexity. The optimal distribution function 
$\mu(\rho)$ is determined from a variational principle, $(\delta
\log Z)/(\delta \mu(\rho))=0$, where $Z$ is computed from 
equ.~(\ref{bound}). One can show that the variational result
for the free energy $F=-\log(Z)/V$ provides an upper bound on 
the true free energy.  

\begin{figure}
\centering
\includegraphics[width=14cm]{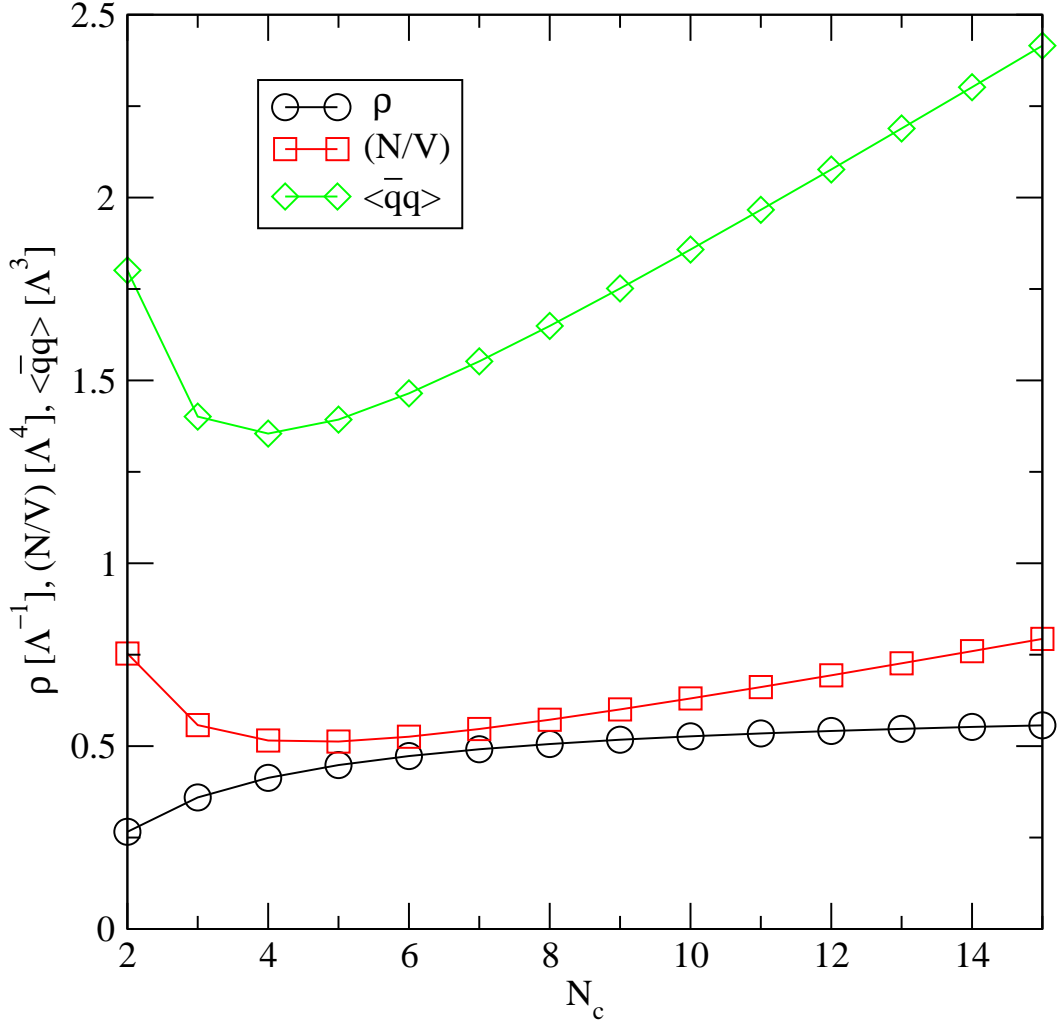}  
\caption{\label{fig_mfa}
Average instanton size $\rho$, instanton density $(N/V)$ and 
quark condensate $\langle\bar{q}q\rangle$ for different numbers
of colors $N_c$. The results shown in this figure were obtained 
using the mean field approximation. All quantities are given in 
units of the QCD scale parameter $\Lambda$. }
\end{figure}

   The calculation of $\langle S-S_1 \rangle$ reduces to the 
calculation of the average instanton interaction $\langle S_{int}
\rangle$. Since the variational ansatz does not include any 
correlations, we only need to average the instanton interaction 
over the collective coordinates of the two instantons. The 
dipole interaction (\ref{int_dip}) vanishes when averaged over
all color orientations. In \cite{Diakonov:1983hh} it was
proposed to compute the instanton interaction at all distances
using a specific ansatz (called the ``sum ansatz'') for the
two-instanton configuration. The result is that both the 
instanton-instanton ($II$) and instanton-anti-instanton ($IA$) 
are repulsive on the average. We find 
\bea
\label{av_int}
 \langle S_{int}\rangle  &=& \frac{8\pi^2}{g^2}
  \gamma^2\rho_I^2\rho_J^2, \hspace{1cm}
  \gamma^2 = \frac{27}{4}\frac{N_c}{N_c^2-1}\pi^2.
\eea 
The interaction contains an explicit factor $N_c/(N_c^2-1)
\sim 1/N_c$ which reflects the probability that two random 
instantons overlap in color space. Since the classical action 
scales as $S_0\sim 1/g^2$ we find that the average interaction
between any two instantons is $O(1)$. Applying the variational 
principle, one finds \cite{Diakonov:1983hh}
\bea
\label{reg_dis}
\mu(\rho) &=& n(\rho)\exp\left[ -\beta\gamma^2
 \left(\frac{N\overline{\rho^2}}{V}\right)\rho^2\right],
\eea
where $\beta=\beta(\overline{\rho})$ is the average instanton
action and $\overline{\rho^2}$ is the average size. We observe
that the single instanton distribution is cut off at large sizes
by the average instanton repulsion. The instanton density and
average size are given by
\bea
\label{dens_mfa}
\frac{N}{V} &=& \Lambda^4 \left[ C_{N_c}\beta^{2N_c} \Gamma(\nu)
(\beta\nu\gamma^2)^{-\nu/2}\right]^{\frac{2}{2+\nu}},\\
\label{rho_mfa}
\overline{\rho^2} &=& \left(\frac{\nu V}{\beta\gamma^2 N}\right)^{1/2},
\hspace{1cm}\nu = \frac{b-4}{2}.
\eea
We note that $\Lambda$ is the only dimensionful parameter. The 
free energy is given by
\be
\label{eps_MFA}
 F = -\frac{b}{4} \left(\frac{N}{V}\right),
\ee
which is in agreement with the trace anomaly. We can now study the 
dependence of $(N/V)$ and $\rho$ on $N_c$, see Fig.~\ref{fig_mfa}. 
We note that to one-loop order the scale in the pre-exponent $\beta
(\bar{\rho})$ is not well determined. In practice we assume that 
$\beta=N_cs_0$ with $s_0=5$. Changing $s_0$ does affect both 
$(N/V)$ and $\rho$ but the main effect can be absorbed in the 
scale parameter. The remaining dependence on $s_0$ is very 
weak. 

  Fig.~\ref{fig_mfa} shows that for $N_c>4$ the average instanton
size is essentially constant while the instanton density grows
linearly with $N_c$. This is easily verified by inspecting 
equ.~(\ref{dens_mfa}). Expanding $\log(N/V)$ in powers of $N_c$
and $\log(N_c)$ we observe that independent of the details
of the interaction the instanton density scales at most as 
a power, not an exponential, in $N_c$ \cite{Ilgenfritz:1980bm}. 
Using the fact that $\gamma^2=O(1/N_c)$, which is equivalent 
to $\langle S_{int} \rangle = O(1)$, we find that $(N/V)=O(N_c)$. 
This result depends on the instanton interaction, but as we noted 
above, $\gamma^2=O(1/N_c)$ is a consequence of the fact that 
instantons are $SU(2)$ gauge field configurations.

  There is a simple argument that explains why the instanton
density scales as the number of colors. In our model, the 
size distribution is regularized by the interaction between
instantons. This means that there has to be a balance between
the average single instanton action and the average interaction
between instantons. If the average instanton action satisfies
$S_0=O(N_c)$ we expect that $\langle S^{tot}_{int}\rangle =O(N_c)$ 
also. Using $\langle S^{tot}_{int}\rangle = (N/V) \langle S_{int} 
\rangle$ and the fact that the average interaction between 
any two instantons satisfies $\langle S_{int} \rangle = O(1)$
we expect that the density grows as $N_c$. 

  Fig.~\ref{fig_mfa_rho} shows the instanton size distribution
for different numbers of colors. We observe that the number 
of small instantons is strongly suppressed as $N_c\to\infty$ 
but the average size stabilizes at a finite value $\bar{\rho}
<\Lambda^{-1}$. We also note that there is critical size
$\rho^*$ for which the number of instantons does not change 
as $N_c\to\infty$. The value of $\rho^*$ is easy to determine
analytically. We write $n(\rho)=\exp(N_cF(\rho))$ with $F(\rho)
=a\log(\rho)+b\rho^2+c$ where the coefficients $a,b,c$ are 
independent of $N_c$ in the large $N_c$ limit. The critical
value of $\rho$ is given by the zero of $F(\rho)$. We find
$\rho^*=0.49\Lambda^{-1}$. The existence of a critical
instanton size for which $n(\rho^*)$ is independent of $N_c$ 
was discussed by \cite{Teper:1979tq,Neuberger:1980as,Shuryak:1995pv}.
The problem was studied on the lattice by Lucini and Teper
\cite{Lucini:2001ej}, who find $\rho^*= 6a = 0.43$ fm.

  Next we wish to study fluctuations in the instanton liquid. 
Fluctuations in the net instanton number are related to the 
second derivative of the free energy with respect to $N$. We 
find
\be
\label{n_fluc}
 \langle N^2\rangle -\langle N\rangle ^2
 =\frac{4}{b}\langle N\rangle .
\ee
This result is in agreement with a low energy theorem based 
on broken scale invariance \cite{Novikov:xj}
\be
\label{scal_let}
\frac{1}{(32\pi^2)^2} \int d^4x\; 
 \left\{ \langle g^2G^2(0)g^2G^2(x)\rangle
 -\langle g^2G^2(0) \rangle^2 \right\}  
 = \frac{4}{b}\frac{1}{32\pi^2}\langle g^2G^2\rangle .
\ee
This result is very general and based solely on the 
renormalization group equations. The left hand side is 
given by an integral over the field strength correlator, 
suitably regularized and with the constant term $\langle 
G^2\rangle^2$ subtracted. For a dilute system of instantons 
equ.~(\ref{scal_let}) reduces to equ.~(\ref{n_fluc}).
The result (\ref{n_fluc}) shows that fluctuations of 
the instanton ensemble are suppressed by $1/N_c$. This
is agreement with general arguments showing that  
fluctuations are suppressed in the large $N_c$ limit. We 
also note that the result (\ref{n_fluc}) clearly shows that 
even if instantons are semi-classical, interactions between 
instantons are crucial in the large $N_c$ limit.

 Fluctuation in the topological charge can be studied 
by adding a $\theta$-term to the partition function
(\ref{Z}). We find
\be
\label{q2_mfa}
 \langle Q^2 \rangle = \langle N \rangle ,
\ee
which is identical to the result in the random instanton 
liquid and not in agreement with Witten's hypothesis
$\chi_{top}=O(1)$. However, Diakonov et al.~noticed
that equ.~(\ref{q2_mfa}) is a consequence of the fact 
that in the sum ansatz the average interaction between
instantons of the same charge is identical to the average
interaction between instantons of opposite charge 
\cite{Diakonov:1995qy}. In general there is no reason for 
this to be the case and more sophisticated instanton 
interactions do not have this feature 
\cite{Shuryak:1988rf,Khoze:1991mx,Verbaarschot:1991sq}.
If $r$ denotes the ratio of the average interaction between
instantons of opposite charge and instanton of the same
charge, $r=\langle S_{IA}\rangle/\langle S_{II}\rangle$, 
then \cite{Diakonov:1995qy}
\be
 \langle Q^2 \rangle = \frac{4}{b-r(b-4)}\langle N\rangle.
\ee
This result shows that for any value of $r\neq 1$ fluctuations
in the topological charge are suppressed as $N_c\to\infty$.
We also note that $\chi_{top}=O(1)$, in agreement with 
Witten's hypothesis. 

\section{Chiral Symmetry Breaking}
\label{sec_csb}

  In this section we wish to study chiral symmetry breaking
in the mean field approximation. This can be done by studying
the Dyson-Schwinger equation for the quark propagator or 
by analyzing the spectrum of the Dirac operator. In this
section we wish to use the more microscopic approach and
analyze the spectrum of the Dirac operator. In a basis
spanned by the individual zero modes of the instantons and
anti-instantons the Dirac operator has the structure
\be
 \left( i\Dslash \right) =
 \left( \begin{array}{cc} 
 0 & T_{IA} \\ T_{IA}^\dagger &0 \end{array}
 \right),
\ee
where $T_{IA}$ is the overlap matrix element of the Dirac
operators between an instanton and anti-instanton 
zero mode. The matrix elements depend on the collective 
coordinates of the instanton. If the interaction between 
instantons is weak, the matrix elements are distributed 
randomly with zero average, but the second moment of $T_{IA}$
is non-zero. Averaging over the positions and orientations 
of the instantons we get 
\be
\label{TIA_var}
\langle |T_{IA}^2|\rangle = \frac{2\pi^2}{3N_c}\frac{N \rho^2}{V}.
\ee
The factor $1/N_c$ comes from the average over $SU(N_c)$. 
Equ.~(\ref{TIA_var}) implies that the average matrix element
of the Dirac operators decreases as $1/N_c$ but the second
moment in the zero modes zone is $O(1)$. If the matrix elements 
are distributed according to a Gaussian unitary ensemble, the 
spectral density is a semi-circle
\be
\label{semi_circ}
\rho(\lambda) = \frac{N}{\pi\sigma V} \left( 1 - 
\frac{\lambda^2}{4\sigma^2} \right)^{1/2},
\ee
with $\sigma^2=|T_{IA}^2|$. We observe that the width of the 
zero mode zone is related to $\sigma$, which is $O(1)$. According 
to the Banks-Casher formula the quark condensate is related to the 
spectral density at zero virtuality
\be
\langle \bar qq\rangle  = -\frac{1}{\pi\rho} \left( \frac{3N_c}{2}
\frac{N}{V}\right)^{1/2}. 
\ee 
Because $(N/V)=O(N_c)$ the quark condensate also grows as $N_c$. 
This, of course, is the expected behavior. We note, however, that
the linear growth in $N_c$ is really a combination of two effects.
The linear increase in the number of modes $N$ provides one factor 
of $N_c^{1/2}$ and the decrease in the average matrix element
$|T_{IA}|$ contributes another factor of $N_c^{1/2}$.

\begin{figure}
\centering
\includegraphics[width=14cm]{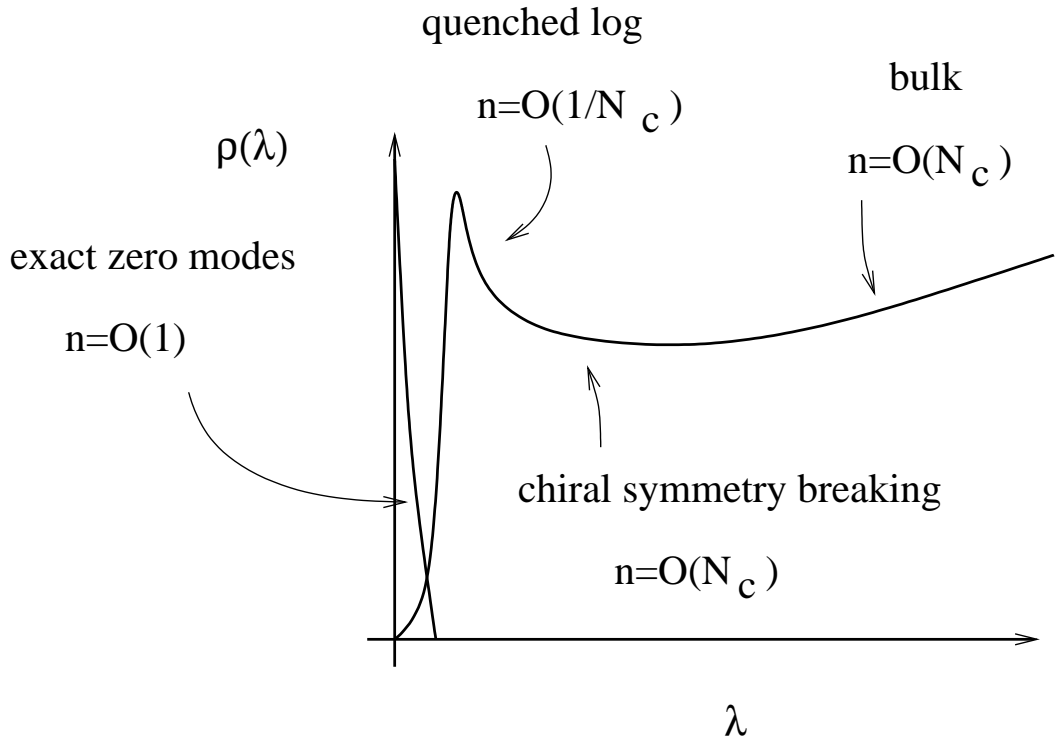}  
\caption{\label{fig_spec_schem}
Schematic behavior of the spectrum of the Dirac operator 
in quenched QCD with many colors.}
\end{figure}

 The true spectral density of the Dirac operator in not given
by a semi-circle. Schematically, the spectrum of the Dirac 
operator in quenched QCD is shown in Fig.~\ref{fig_spec_schem}.
There are several notable features. First, in a finite volume
there is a certain number of exact zero modes. This number
is proportional to $(\chi_{top} V)^{1/2}$ and therefore scales
as $N_c^0$. We note that in the infinite volume limit  
exact zero modes are not important because their number scales 
as $\sqrt{V}$ while the total number of states increases
linearly with the volume. The second feature of the spectrum
is the logarithmic enhancement of the spectrum at small 
virtuality. This enhancement is an artifact of the quenched
approximation. From quenched chiral perturbation theory we 
expect \cite{Toublan:1999hi}
\be
\label{qu_log}
 \rho(\lambda) = \frac{\Sigma}{\pi} 
 \left\{ 1 - \frac{m_0^2}{16\pi^2f_\pi^2}
   \log\left(\frac{|\lambda|}{\mu}\right) + \ldots \right\}.
\ee
Here, $\Sigma$ is a parameter that corresponds to the 
(negative) unquenched chiral condensate and $m_0$ is the 
mass of the quenched ghost pole. This mass corresponds to 
the mass of the $\eta'$ in the chiral limit of full QCD.
Using $m_0^2=O(1/N_c)$ we find that the coefficient of the 
logarithmic enhancement vanishes as $1/N_c$ in the large 
$N_c$ limit. This, of course, is consistent with the idea
that the fermion determinant is not important in the large
$N_c$ limit. The third component of the spectrum is given
by the almost zero modes related to chiral symmetry breaking. 
This part of the spectrum is expected to scale as $N_c$, 
as is the bulk of the spectrum which is not related to
chiral symmetry breaking. 

 At finite $N_c$ the spectrum of the Dirac operator in full
QCD with light fermions has a different behavior. The number of 
exact zero modes is again proportional to $(\chi_{top}V)^{1/2}$ 
but in full QCD the topological susceptibility is suppressed, 
$\chi_{top}\simeq -m\langle \bar{\psi}\psi\rangle$. This 
implies that as $N_c$ increases the number of exact zero
modes initially increases as $N_c^{1/2}$ and then saturates
at the value corresponding to the quenched topological 
susceptibility. In the infinite volume limit the spectrum 
near the origin is linear. Chiral perturbation theory 
predicts the slope of the spectrum \cite{Smilga:1993in}
\be
 \rho(\lambda) = \frac{\Sigma}{\pi}\left\{
 1 + \frac{(N_f^2-4)\Sigma}{32\pi N_f f_\pi^4}|\lambda|
  + \ldots \right\}.
\ee
We observe that the constant part of the spectrum grows 
as $N_c$ whereas the linear part is independent of $N_c$. 
This implies that in the large $N_c$ limit the spectrum 
at the origin is flat for any number of flavors as long 
as $N_f$ is not of the order $N_c$.

\section{The Interacting Instanton Liquid}
\label{sec_ilm}

 In this section we wish to go beyond the mean field approximation
and study the partition function of the instanton liquid using 
numerical simulations. These simulations take into account 
all correlations between instantons. The numerical techniques
are described in detail in \cite{Shuryak:1994rr,Schafer:1995pz}. 
In order to perform these simulations we have to fully
specify the instanton interaction. We have used the ``streamline'' 
interaction determined in \cite{Khoze:1991mx,Verbaarschot:1991sq}.
The streamline solution is characterized by the fact that
the action of the approximate instanton-anti-instanton
solution is a local minimum except in the direction of
the ``valley'' in configuration space that connects 
a well separated IA pair with a very close pair. There 
is no interaction between two instantons of the same charge.
The interaction between two instantons of opposite charge 
approaches $-2S_0$ if the relative color orientation is attractive. 

  This implies that the streamline interaction lacks the 
repulsive core that is required to stabilize the instanton 
ensemble at the classical level. In order to correct this
problem we have added a purely phenomenological core 
to the instanton interaction. The interaction is given 
by 
\be
\label{core}
   S_{\rm core} = \frac{8\pi^2}{g^2} \frac{A}{\lambda^4}
   |u|^2, \hspace{1cm}
 \lambda = \frac{R^2+\rho_I^2+\rho_A^2}{2\rho_I\rho_A}
  + \left( \frac{(R^2+\rho_I^2+\rho_A^2)^2}{4\rho_I^2\rho_A^2}
   - 1\right)^{1/2}
\ee
in both $II$ and $IA$ channels. The dimensionless parameter 
$A$ controls the strength of the core and is adjusted to 
reproduce the phenomenological diluteness $\rho^4 (N/V)$ of 
the instanton ensemble. In our simulations we have used
$A=128$. We note that the hard core interaction
equ.~(\ref{core}) only acts between instantons that overlap
in color space. As a result, we expect $\langle S_{core}
\rangle = O(1)$, in agreement with the interaction used 
in the mean field treatment. 

 One might argue that there should not be any interaction
between instantons of the same charge because there 
is a $2\times(2N_c)$ parameter family of two-instanton
solutions with action $2\times (8\pi^2)/g^2$ \cite{Atiyah:ri}. 
However, there are two phenomena that lead to an effective 
instanton-instanton interaction. The first is related to the 
fact that the collective coordinate measure for two close 
instantons is not just the product of two single instantons 
measures. Carter and Shuryak argued that this will lead
to an effective short range instanton-instanton repulsion
\cite{Carter:2001ih,GarciaPerez:1999zk}. Whether this effect 
has the same $N_c$ dependence as the classical interaction 
equ.~(\ref{core}) is an important problem. The second effect 
is that quantum corrections to a charge two instanton solution 
do not factorize. As a quantum correction, this effect is 
naively suppressed by $1/S_0\sim 1/N_c$, but the suppression 
can be overcome if instantons with different color orientation 
interact.

  In order to determine the instanton density and the free
energy of the instanton liquid we have to compute the partition
function equ.~(\ref{Z}). Monte Carlo simulations are ideally
suited for computing expectation values, but they do not 
directly provide the partition function. It is possible, 
however, to compute free energy differences. This implies
that Monte Carlo simulations can be used to compute the 
ratio of two partition functions. In practice we calculate
the ratio of the exact partition function and the variational
estimate equ.~(\ref{Z_var}). The exact partition function
is given by \cite{Schafer:1995pz}
\be
\label{int_coup}
  \log Z = \log (Z_1)
  - \int_0^1 d\alpha\,  \langle  \left( S-S_1\right)
     \rangle_{\alpha}.
\ee
Here, $Z_1$ is the variational partition function equ.~(\ref{Z_var})
and $S_1$ is the variational action. The expectation value 
$\langle .\rangle_\alpha$ is determined using the interpolating
action $S_\alpha=S_1+\alpha(S-S_1)$. $S_\alpha$ reduces to the 
variational action for $\alpha=0$ and the exact action for 
$\alpha=1$. 

\begin{figure}
\centering
\includegraphics[width=14cm]{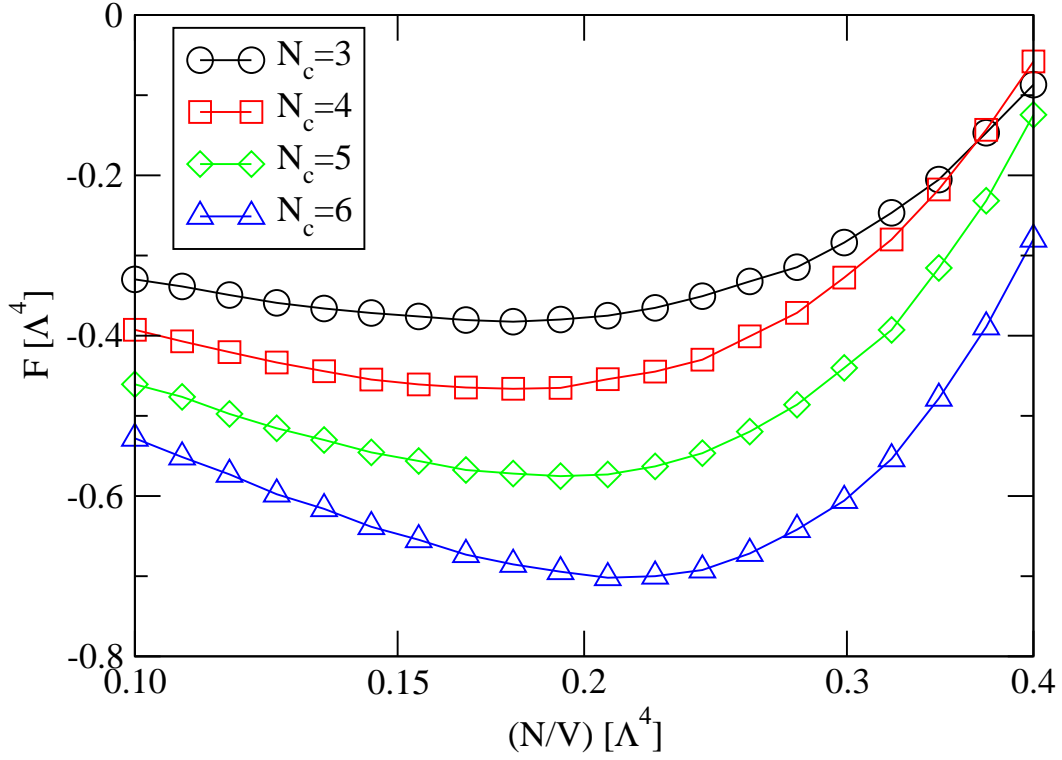}  
\caption{\label{fig_f}
Free energy $F$ of the quenched instanton liquid as a function of 
the instanton density $(N/V)$ for $N_c=3,\ldots,6$ colors. Both
$(N/V)$ and $F$ are given in units of the QCD scale parameter.  
The results shown in this figure were obtained using numerical
simulations with $N=32$ instantons.}
\end{figure}

\begin{figure}
\centering
\includegraphics[width=14cm]{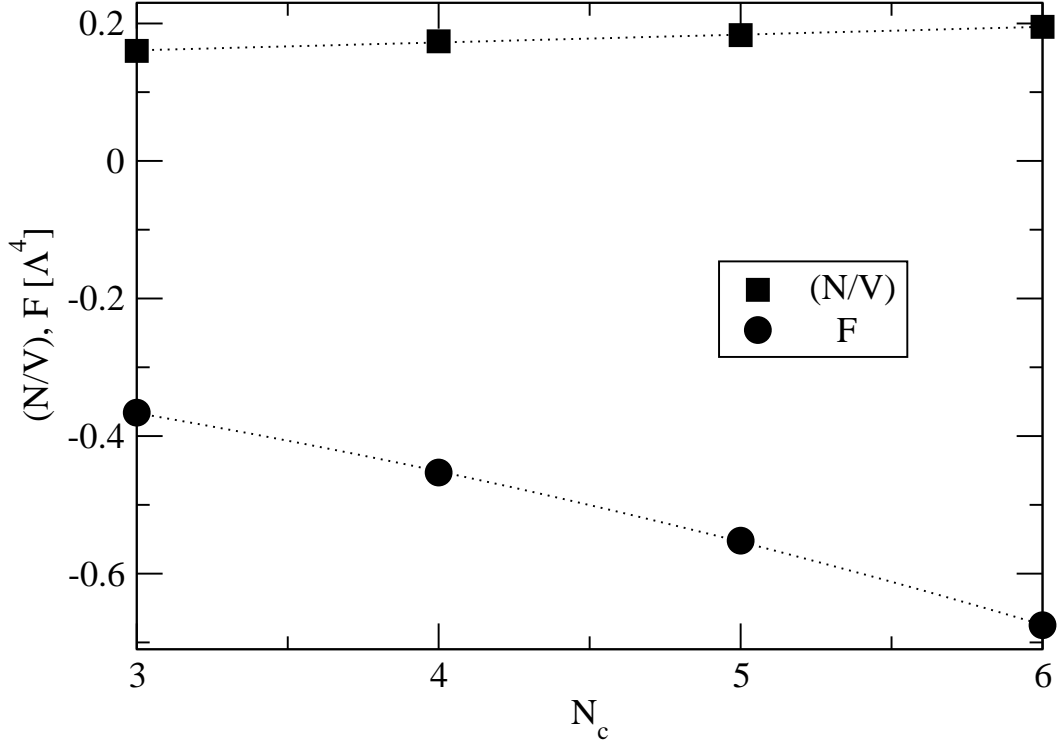}  
\caption{\label{fig_dens}
Instanton density $(N/V)$ and free energy $F$ in a pure 
gauge instanton ensemble for $N_c=3,\ldots,6$ colors. 
Both $(N/V)$ and $F$ are given in units of $\Lambda^4$
where $\Lambda$ is the QCD scale parameter. The dashed
lines show fits of the form $a_1N_c^2+a_2N_c+a_3$ (for
the free energy $F$) and $a_2N_c+a_3$ (for the instanton
density $N/V$). }
\end{figure}

  The free energy energy of the instanton liquid as a function
of the instanton density for $N_c=3,\ldots,6$ is shown in 
Fig.~\ref{fig_f}. The equilibrium density is determined by
the minimum of the function $F(N/V)$. The dependence of
equilibrium density and the free energy on the number of
colors is shown in Fig.~\ref{fig_dens}. We observe that 
$(N/V)$ increases linearly with $N_c$ whereas the free
energy is quadratic. The slope of $(N/V)$ as a function 
of $N_c$ is small, in agreement with the mean field result
shown in Fig.~\ref{fig_mfa}. In contrast to the mean field result 
the linear behavior already sets in at small $N_c\simeq 3$.
We have checked the stability of our results to changing
the strength of the hard core interaction and including higher
order corrections in the QCD beta function. Both changes affect
the results quantitatively but not qualitatively. However,
the results are crucially dependent on the assumption that
the parameter $A$ in equ.~(\ref{core}) is not a function
of $N_c$. 

 We have also studied the instanton size distribution,
the topological susceptibility and the spectrum of the 
Dirac operator for different numbers of colors. In order 
to be able to distinguish more clearly between different
scenarios we have not used the exact instanton density 
determined in Fig.~\ref{fig_f} but have simply scaled
$(N/V)\sim N_c$. At large $N_c$ this will only introduce
errors that are suppressed by $1/N_c$. The instanton
size distribution is shown in Fig.~\ref{fig_nrho}. As 
expected small instantons are strongly suppressed as
the number of colors increases. We observe a clear fixed 
point in the size distribution at $\rho^*\Lambda \simeq 
0.27$.

\begin{figure}
\centering
\vspace*{3cm}
\includegraphics[width=14cm]{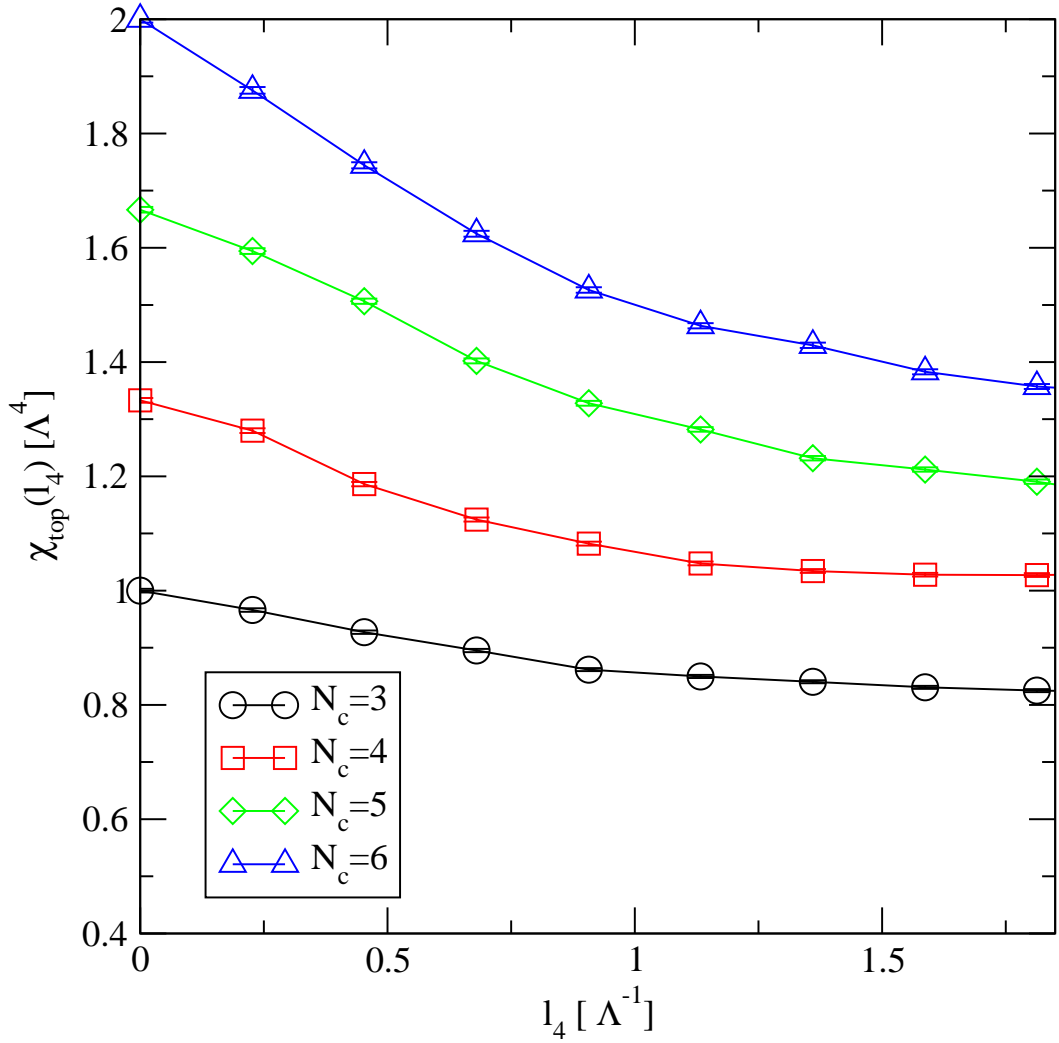} 
\caption{\label{fig_suscep}
Finite volume topological susceptibility $\chi_{top}(l_4)$
in a pure gauge instanton ensemble for different numbers of 
colors. The results were obtained using numerical simulations
with $N=128$ instantons.}
\end{figure}

\begin{figure}
\centering
\includegraphics[width=14cm]{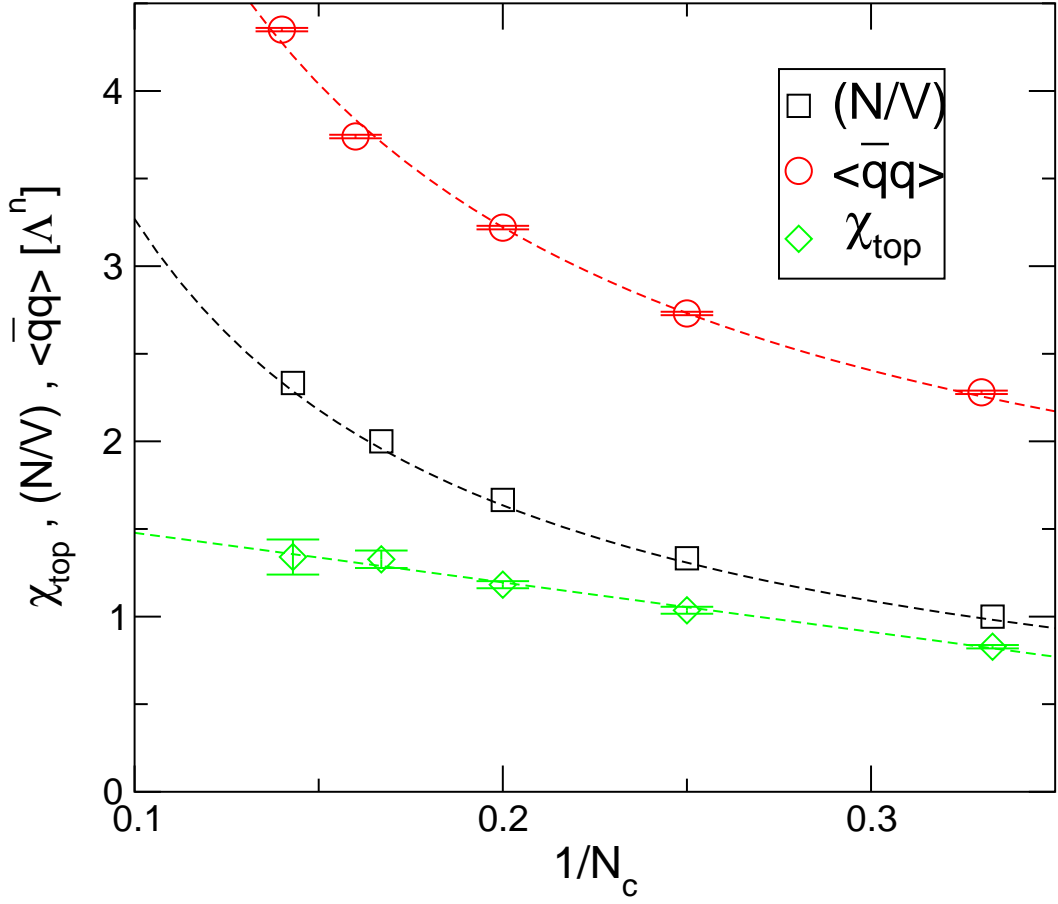}  
\caption{\label{fig_obs}
Dependence of the chiral condensate $\langle\bar{\psi}\psi\rangle$
and the topological susceptibility $\chi_{top}$ on the number
of colors. The instanton density $(N/V)$ was assumed to scale 
as $(N/V)\sim N_c$. The dashed lines show fits of the form 
$a_1N_c+a_2$ (for $\langle\bar{\psi}\psi\rangle$ and $N/V$)
and $a_2+a_3/N_c$ (for $\chi_{top}$).}
\end{figure}

 Our simulations were carried out in the total topological charge
$Q_{top}=0$ sector of the theory. We can nevertheless determine
the topological susceptibility by measuring the average $Q^2_{top}$ 
in a sub-volume $V_3\times l_4$ of the euclidean box $V_3\times L_4$ 
\cite{Shuryak:1994rr}. The finite volume susceptibility is given by
\be
\chi_{top}(l_4) = \frac{\langle Q^2_{top} \rangle_{V_3\times l_4}}
  {V_3\times l_4} \left(1-\frac{l_4}{L_4}\right)^{-1}.
\ee
The factor $(1-l_4/L_4)^{-1}$ takes into account the constraint 
from overall charge neutrality. This correction factor is derived
under the assumption that the fluctuations are Gaussian. In an 
ideal calculation $L_4\gg l_4$ and the correction for overall 
neutrality is small. The topological susceptibilities are shown 
in Fig.~\ref{fig_suscep}. We observe that $\chi_{top}(l_4)$
tends to a constant as $l_4$ increases. We identify this 
constant with the susceptibility in the thermodynamic limit. 
We find that for $N_c=3$ the topological susceptibility agrees
well with the expectation based on Poissonian statistics, 
$\chi_{top}\simeq (N/V)$. For $N_c>3$, however, fluctuations
are significantly suppressed and the topological susceptibility 
increases more slowly than the density of instantons. 
Fig.~\ref{fig_obs} shows that our results are consistent
with a scenario in which $\chi_{top}$ remains finite as 
$N_c\to\infty$. 

\begin{figure}
\centering
\vspace*{3cm}
\includegraphics[width=14cm]{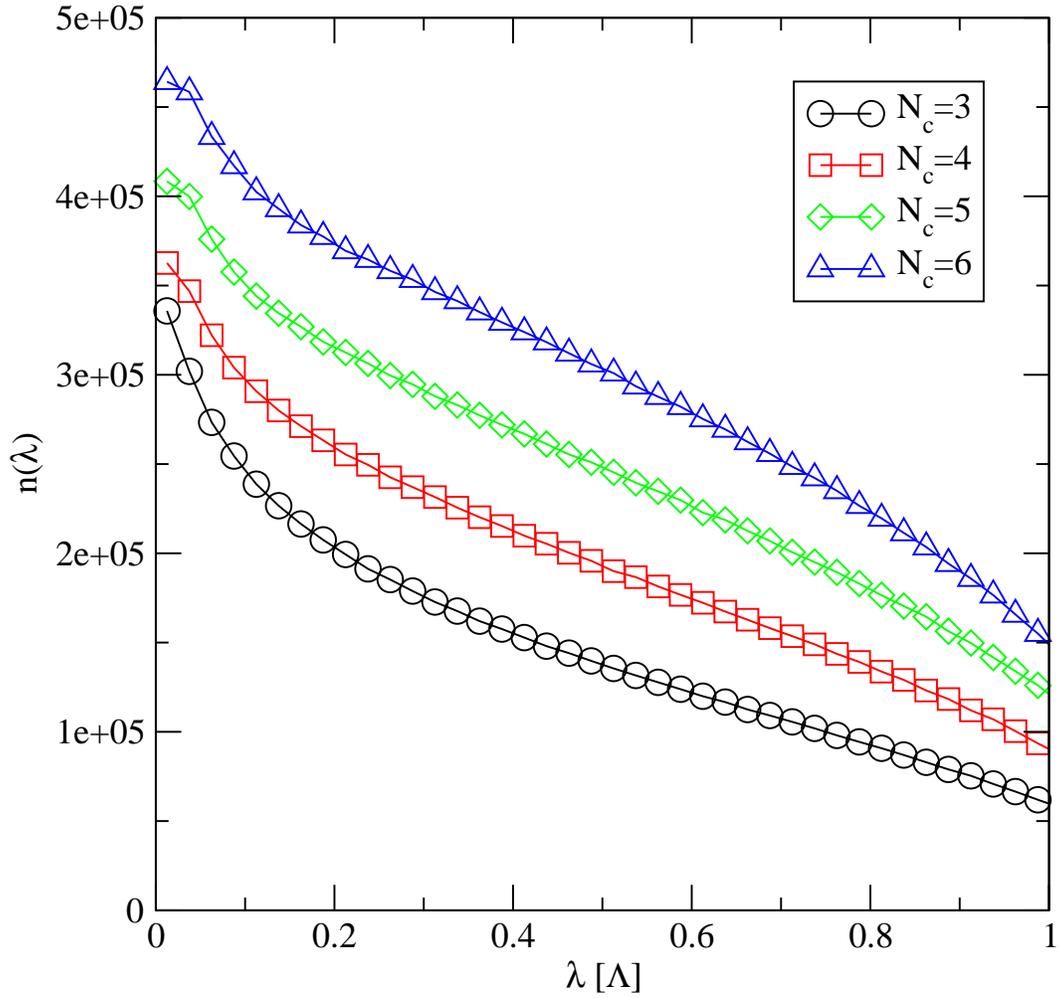} 
\caption{\label{fig_dirac}
Spectrum of the Dirac operator in a pure gauge instanton 
ensemble for different numbers of colors. The eigenvalues
are given in units of the QCD scale parameter. The results were
obtained using numerical simulations with $N=128$ instantons. }
\end{figure}

  In Fig.~\ref{fig_dirac} we show the spectrum of the Dirac
operator for $N_c=3,\ldots,6$. Since we calculate in the 
$Q_{top}=0$ sector of the theory there are no exact zero modes.
We clearly observe the enhancement of the spectral density 
near $\lambda=0$, but we also note that this enhancement 
becomes weaker as the number of colors increases. The chiral
condensate for $m_q=0.1\Lambda$ is shown in Fig.~\ref{fig_obs}.
We clearly see that $\langle\bar{q}q\rangle$ is linear in $N_c$.

 In the recent literature a number of authors have studied
the role of instantons in chiral symmetry breaking by analyzing
the local chirality $X(x)$ of low lying eigenstates of the 
Dirac operator \cite{Horvath:2001ir}. The quantity $X(x)$
is defined by
\be
 \tan\left(\frac{\pi}{4}\left( 1+X(x) \right) \right) =
  \left( \frac{\psi^\dagger(1+\gamma_5)\psi}
              {\psi^\dagger(1-\gamma_5)\psi}\right)^{1/2},
\ee
where $\psi$ is an eigenfunction of the Dirac operator
$i\Dslash\psi=\lambda \psi$ in a given gauge configuration. 
In order to study chiral symmetry breaking one only considers
the lowest few eigenvectors. The instanton liquid model 
predicts that these eigenvectors are linear combinations 
of instanton and anti-instanton zero modes. This implies
that at points $x$ where the wave function $\psi^\dagger\psi$ 
is large it is either left or right handed, $X(x)\simeq\pm 1$.
In order to test this prediction one has to choose a cutoff
on the eigenvalue $\lambda$ and some cutoff on the magnitude
of $\psi^\dagger\psi$. Typically, the points $x$ are restricted
to the top few percent of the eigenfunction. The instanton
model suggests that this fraction should be no larger than 
the diluteness of the instanton liquid, and that the maximum
eigenvalue should be smaller than the width $|T_{IA}|$ of the 
zero mode zone. 

  In Fig.~\ref{fig_isgur} we show numerical results for the 
local chirality distribution in the instanton liquid model. 
We have included all states in the zero mode zone and used
the top 30\% of the wavefunction. The $N_c$ dependence is
similar if more restrictive cuts are used. We observe that 
the double peak structure is very pronounced for all $N_c=3,
\ldots, 6$. There is a small shift of the peaks toward
smaller values of $X$ as the number of colors increases.
We have verified that for $N_c>10$ the double peak structure
disappears. This is related to the fact that the instanton
density increases and instantons overlap in space. The 
instanton liquid remains dilute, however, because instantons
do not overlap in color space. This is easily verified by
computing chirality distributions for eigenstates of the 
Dirac operator projected on a $SU(2)$ subgroup. Cundy et 
al.~computed local chirality distributions in lattice 
QCD for $N_c=2,\ldots,5$ \cite{Cundy:2002hv}. In their 
calculations the double peak structure disappears much more 
quickly as the number of colors increases. This implies that 
either the instanton ensemble is not as dilute as we have
assumed, or that mixing with non-zero modes is more important. 
Given the fact the value of $\rho^*$ in our simulations is
also smaller than the value obtained by Lucini and Teper,
it seems likely that the instanton liquid in QCD is not
quite as dilute as suggested in \cite{Shuryak:1981ff}.

\section{Hadronic Correlation Functions}
\label{sec_cor}

 We have also studied hadronic correlation functions in the
instanton liquid for different numbers of colors. We have
considered, in particular, correlation functions of currents
with the quantum numbers of the pion, the rho meson, and
the $\eta'$ meson. The currents are given by
\be
 j_\pi = \bar{d}i\gamma_5 u, \hspace{1cm}
 j_\rho^\mu = \bar{d}\gamma^\mu u, \hspace{1cm}
 j_{\eta'} = \frac{1}{\sqrt{2}} 
  ( \bar{u}i\gamma_5u+\bar{d}i\gamma_5 d).
\ee
The pion and rho meson are flavor non-singlet mesons. The
corresponding correlation functions only involve the 
connected contribution
\be
\label{mes_cor}
 \Pi_{I=1}(x) = \langle {\rm Tr}\left[ S^{ab}(0,x)\Gamma
 S^{ba}(x,0)\Gamma \right] \rangle ,
\ee
with $\Gamma=i\gamma_5,\gamma_\mu$ for the pion and rho 
meson, respectively. Here, $S^{ab}$ denotes the quark 
propagator in a given gauge configuration, $a,b$ are 
color indices, and the trace runs over Dirac indices.
The average $\langle .\rangle$ is performed with respect
to the partition function equ.~(\ref{Z}). The $\eta'$ meson 
is a flavor singlet meson and the correlation function has 
an additional, disconnected, term. We find
\be
\label{sing_cor}
 \Pi_{I=0}(x) = \langle {\rm Tr}\left[ S^{ab}(0,x)\Gamma
 S^{ba}(x,0)\Gamma \right] \rangle  - 
 2 \langle {\rm Tr}\left[ S^{aa}(0,0)\Gamma
 \right]\, {\rm Tr}\left[ S^{bb}(x,x)\Gamma \right]\rangle ,
\ee
with $\Gamma=i\gamma_5$. The masses of the $\pi,\rho$ and
$\eta'$ mesons can be extracted by fitting the correlation
functions to a spectral representation. At large euclidean
separation the correlation function is determined by the 
lowest hadronic resonance. A good model for the contribution
of excited states is given by the free quark-anti-quark 
continuum above some threshold invariant mass $\sqrt{s_0}$.
The threshold roughly corresponds to the location of the 
first excited state. For the pion and rho meson we use
\bea
\label{spec_pi}
 \Pi_\pi(x) &=& \lambda_\pi^2 D(m_\pi,x) +
  \frac{N_c}{8\pi^2}\int_{s_0}^\infty ds\, 
  s D(\sqrt{s},x), \\
\label{spec_rho}
 \Pi_\rho(x) &=& f_\rho^2 m_\rho^4 D(m_\rho,x) +
  \frac{2N_c}{8\pi^2}\int_{s_0}^\infty ds\, 
  s D(\sqrt{s},x),
\eea
where $D(m,x)$ is the euclidean space propagator for 
a free scalar meson with mass $m$
\be
 D(m,x) = \frac{m}{4\pi^2x}K_1(mx).
\ee
In full QCD the $\eta'$ meson can be described by the 
same spectral representation as the pion. In quenched
QCD, however, the spectral function in the $\eta'$ 
channel is not positive definite. There is a well 
established method for dealing with this problem. In
quenched chiral perturbation theory it is assumed
that the disconnected part of the $\eta'$ correlation
function corresponds to a ghost pole in the spectrum
\cite{Sharpe:1992ft,Bernard:1992mk,Kuramashi:1994aj}.
In momentum space the pole contributions are given by
\be
\label{eta_qu}
\Pi_{\eta'}(q) = \frac{\lambda_\pi^2}{q^2+m_\pi^2}
   -\frac{\lambda_\pi^2}{q^2+m_\pi^2}m_0^2
    \frac{1}{q^2+m_\pi^2}.
\ee
Going from quenched to unquenched QCD corresponds
to summing the geometric series
\be
\Pi_{\eta'}(q)\to \frac{\lambda_\pi^2}{q^2+m_\pi^2+m_0^2}.
\ee
and $m_{\eta'}^2=m_0^2+m_\pi^2$ in full QCD. This result
is expected to be exact in the large $N_c$ limit. In 
addition to that there is evidence from the lattice that the 
mass of the $\eta'$ meson can be extracted from the ghost pole
propagator even for $N_c=3$ \cite{Kuramashi:1994aj}. 
In coordinate space equ.~(\ref{eta_qu}) corresponds to
\be
\label{spec_eta}
 \Pi_\pi(x) = \lambda_\pi^2 D(m_\pi,x) 
  -\lambda_\pi^2 m_0^2 D'(m_\pi,x)
  +\frac{N_c}{8\pi^2}\int_{s_0}^\infty ds\, 
  s D(\sqrt{s},x), \\
\ee
with 
\be
 D'(m,x) = \frac{1}{8\pi^2}K_0(mx).
\ee
We have used the spectral representation equ.~(\ref{spec_eta})
to determine the quenched $\eta'$ mass $m_0$ for different $N_c$.

 Before we come to the numerical studies we would like to 
present several analytical results. If we study correlation
functions at distances small compared to the average separation
between instantons, $x\ll(N/V)^{-1/4}$, it is sufficient
to take into account the contribution from the closest 
instanton only. The quark propagator is given by
\be
\label{S_SIA}
S(x,y) \simeq \frac{\psi_{0}(x-z)\psi^\dagger_{0}(y-z)}{m^*}
\ee
where $\psi_0(x)$ is the zero mode wave function and $z$ is
the location of the instanton. The effect of all other instantons
only enters via the effective mass $m^*\simeq \pi\rho(2/N_c)^{1/2}
(N/V)^{1/2}$. The single instanton contribution to the 
correlation function is found by inserting the propagator 
equ.~(\ref{S_SIA}) into the equations (\ref{mes_cor}) and
(\ref{sing_cor}) and averaging over the collective coordinates
of the instanton. 

\begin{figure}
\centering
\vspace*{3cm}
\includegraphics[width=14cm]{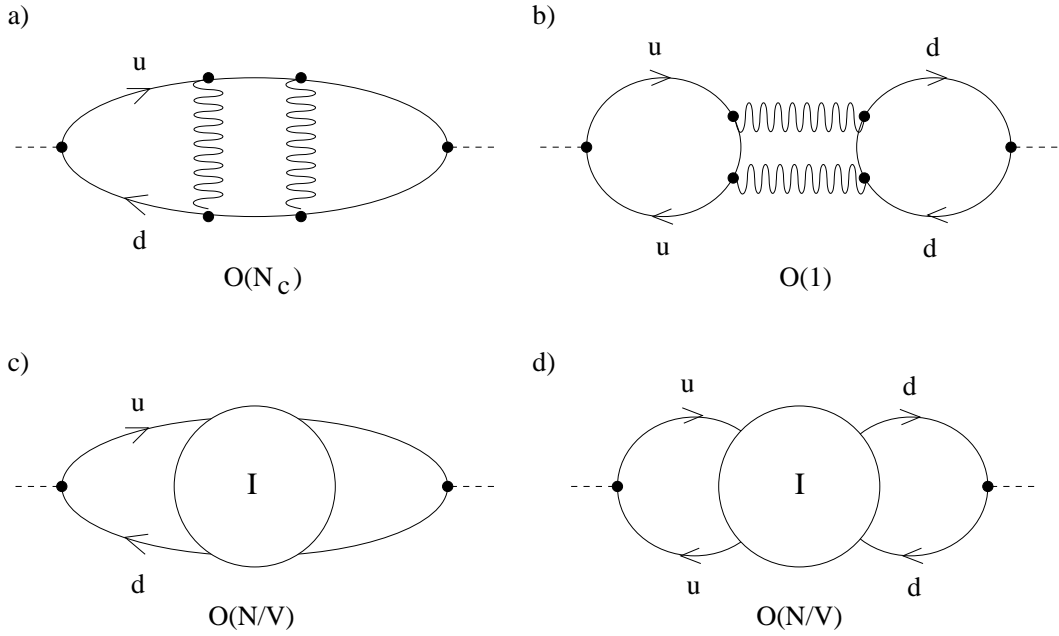} 
\caption{\label{fig_ozi}
Comparison of perturbative and instanton contributions to 
connected and disconnected correlation functions. The perturbative
contribution to the disconnected correlator Fig.~b is suppressed
by a factor $1/N_c$ compared to the connected correlator Fig.~a. 
The single instanton contribution to the two correlation functions, 
shown in Figs.~c and d, is the same, up to a sign.}
\end{figure}

  Because of the chiral structure of the rho meson current 
there is no zero mode contribution to the rho meson correlation 
function. In the pion and $\eta'$ meson channel we find
\cite{Shuryak:1982qx}
\be
\label{ps_SIA}
\Pi^{SIA}_{\pi,\eta'}(x) = \pm\int d\rho\, n(\rho)
 \frac{6\rho^4}{\pi^2}\frac{1}{(m^*)^2}
 \frac{\partial^2}{\partial (x^2)^2}
 \left\{ \frac{4\xi^2}{x^4} \left(
 \frac{\xi^2}{1-\xi^2} +\frac{\xi}{2}
 \log\frac{1+\xi}{1-\xi}\right)\right\},
\ee
with $\xi^2=x^2/(x^2+4\rho^2)$. As is well known, the instanton
contribution is attractive in the pion channel, and repulsive 
in the $\eta'$ channel. The contribution of one instanton 
of given size does not involve any factors of $N_c$, and is 
the same, up to the sign, in the pion and $\eta'$ channel.
This is illustrated in Fig.~\ref{fig_ozi}. Perturbative contributions
to the disconnected correlation function are suppressed by a
factor $1/N_c$ compared to the connected correlator. However,
the single instanton contribution to the disconnected correlator
is exactly the same as the instanton contribution to the 
connected correlator. After integration over the instanton 
distribution the correlation function scales as $N_c$ because 
the instanton density is proportional to $N_c$. This is the 
expected behavior in the case of the pion correlation function 
but it implies that in the single instanton approximation
the anomalous $\pi-\eta'$ splitting does not disappear in the 
large $N_c$ limit. 

   In order to go to large distances, $x>(N/V)^{-1/4}$ we
have to resum the instanton interaction. This can be 
achieved using the mean field (Hartree) and RPA 
methods \cite{Diakonov:1985eg}. The mean field approximation
gives the constituent quark propagator 
\be 
 S_Q(x) = \int\frac{d^4p}{(2\pi)^4} \,e^{ip\cdot x}\;\;
   \frac{\pslash+iM(p)}{p^2+M(p)^2},
\ee
where $M(p)$ is the dynamically generated quark mass. 
The momentum dependence of $M$ is determined by the 
Fourier transform of zero mode wave function. In the 
rho meson channel there is no direct instanton induced
interaction and the correlation function is given 
by two non-interacting constituent quarks. We have
\be
\label{cor_fact}
\Pi_\Gamma(x)^{MFA} = N_c {\rm Tr}\left[ S_Q(x)\Gamma
  S_Q(-x)\Gamma\right] ,
\ee
where the trace is over the Dirac indices only and 
$\Gamma=\gamma_\mu$. In the pion and $\eta'$ channel
there is a direct instanton interaction that can be 
resummed using the RPA. We have 
\cite{Diakonov:1985eg,Hutter:1994av,Kacir:1996qn,Schafer:1996wv}
\be
\label{cor_rpa}
\Pi_{\pi,\eta'}(x)  = \Pi_\pi^{MFA}(x) + \Pi_{\pi,\eta'}^{RPA} \\ 
\ee
with
\be
\label{pscor_int}
\Pi^{RPA}_{\pi,\eta'}(x) = N_c \left(\frac{N_cV}{N}\right)
  \int d^4q\, e^{iq\cdot x}\;\;
           \Gamma_5(q)\frac{\pm 1}{1\mp C_5(q)}\Gamma_5(q) .
\ee
The loop and vertex functions $C_5$ and $\Gamma_5$ are given by
\bea
\label{ps_bubble_mfa}
 C_5 (q) &=& 4N_c\left(\frac{V}{N}\right)
  \int \frac{d^4p}{(2\pi)^4} 
  \frac{M_1M_2(M_1M_2-p_1\cdot p_2)}
       {(M_1^2+p_1^2)(M_2^2+p_2^2)} ,\\
\label{ps_vertex_mfa}
\Gamma_5 (q) &=& \;\; 4  \int \frac{d^4p}{(2\pi)^4} 
  \frac{(M_1M_2)^{1/2}(M_1M_2-p_1\cdot p_2)}
       {(M_1^2+p_1^2)(M_2^2+p_2^2)}  ,
\eea
where $p_1=p+q/2$, $p_2=p-q/2$ and $M_{1,2}=M(p_{1,2})$. 
Using $(N/V)\sim N_c$ and $M\sim 1$ we observe that the 
pion and $\eta'$ correlation functions in the large 
$N_c$ limit  depend on $N_c$ only through an overall
factor of $N_c$. This implies $f_\pi^2=O(N_c)$ and $m_\pi,
m_{\eta'}=O(1)$. 

\begin{figure}
\centering
\vspace*{3cm}
\includegraphics[width=14cm]{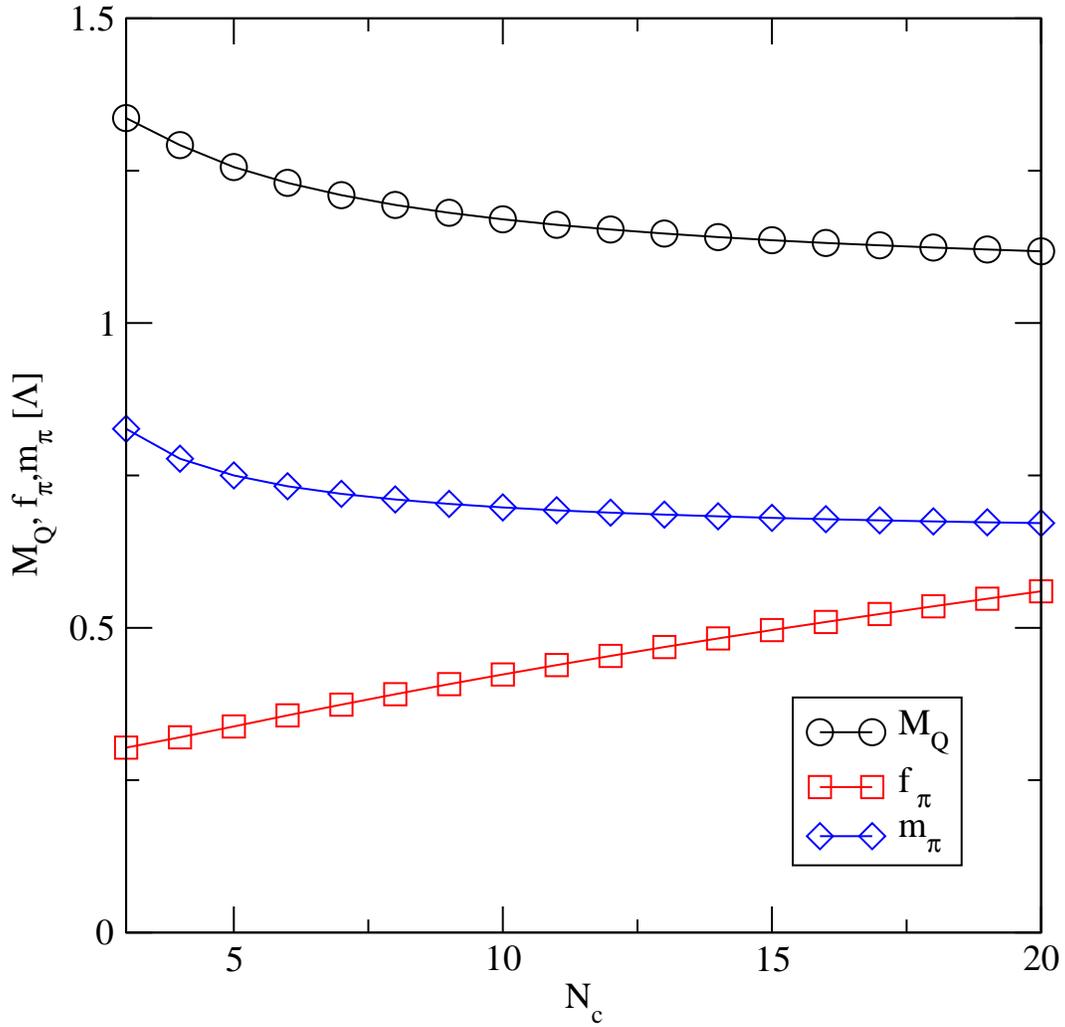} 
\caption{\label{fig_mfa_csb}
Constituent quark mass $M_Q$, pion mass $m_\pi$, and pion decay 
constant $f_\pi$ as a function of the number of colors $N_c$.
All quantities are given in units of the QCD scale parameter 
$\Lambda$. The results shown in this figure were obtained 
using the mean field approximation.}
\end{figure}

  Numerical results for the constituent quark mass $M_Q$
as well as the pion mass $m_\pi$ and pion decay constant
$f_\pi$ are shown in Fig.~\ref{fig_mfa_csb}. We have used
the average instanton size and density determined in
section \ref{sec_mfa}. The current quark mass is $m_q=
0.025\Lambda$. There is some variation in the constituent
quark and pion masses for small $N_c<10$ but the size 
of $1/N_c$ corrections is not large, about 20\% for 
$N_c=3$. The pion decay constant shows the expected 
$N_c^{1/2}$ behavior but in this case $1/N_c$ corrections
are large, about 80\% for $N_c=3$.

\begin{figure}
\centering
\vspace*{3cm}
\includegraphics[width=14cm]{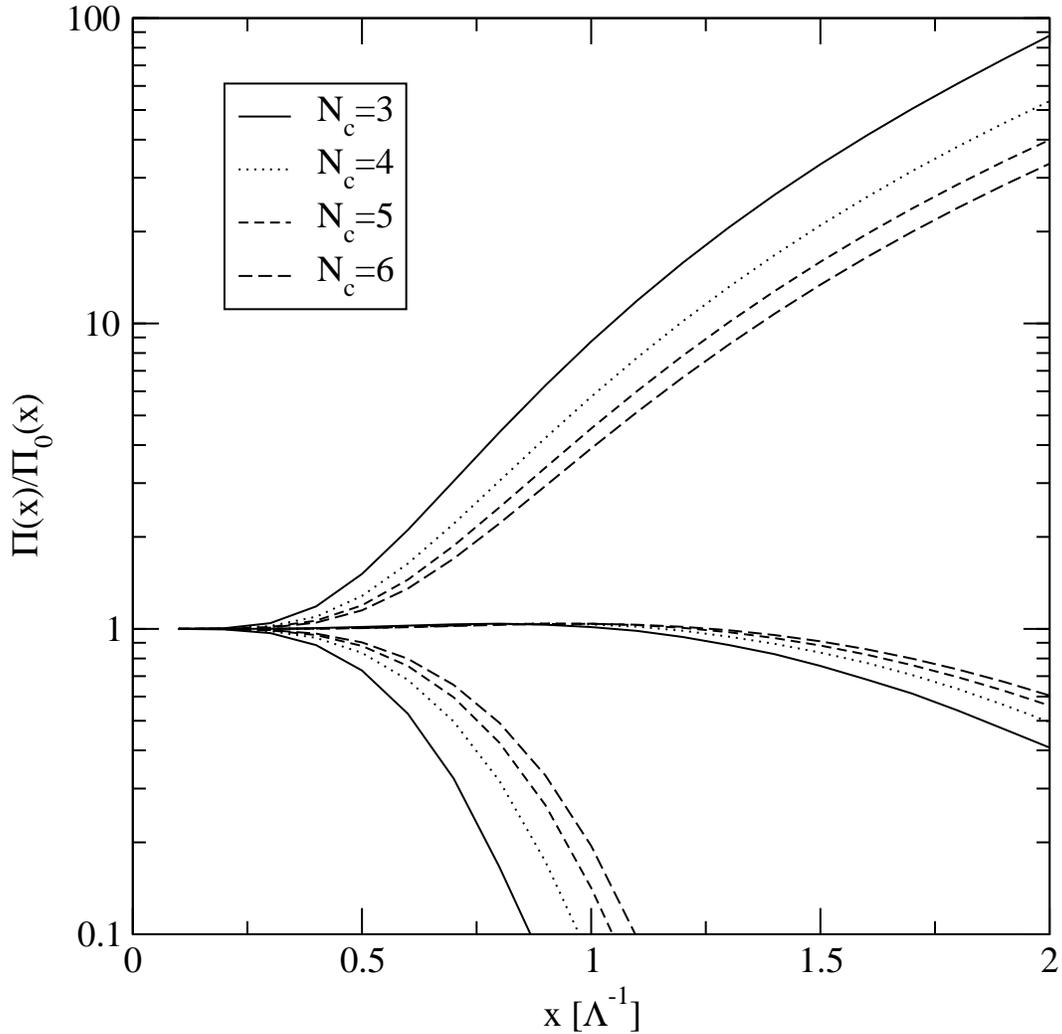} 
\caption{\label{fig_mfa_cor}
Correlation functions in the pion, rho meson, and $\eta'$ meson
channel. The correlators are shown as a function of the distance
in units of the inverse scale parameter. The correlation functions
are normalized to free field behavior, $\Pi_0(x)\sim N_c/x^6$.
The results shown in this figure were obtained using the mean 
field approximation.}
\end{figure}

  In Fig.~\ref{fig_mfa_cor} we show the correlation 
functions in the pion, rho meson, and $\eta'$ meson 
channel. The correlation functions are normalized to
free field behavior. The overall factor of $N_c$ drops
out if the correlator is normalized in this way. We 
observe that the $\rho$ meson correlation function 
is essentially independent of $N_c$ already for small
$N_c$. There are substantial $1/N_c$ corrections in 
the $\pi$ and $\eta'$ channel. The splitting between 
the $\pi$ and $\eta'$ correlation functions is reduced
in going from $N_c=3$ to $N_c=6$ but it remains finite 
and large as $N_c\to\infty$.

\begin{figure}
\centering
\vspace*{3cm}
\includegraphics[width=14cm]{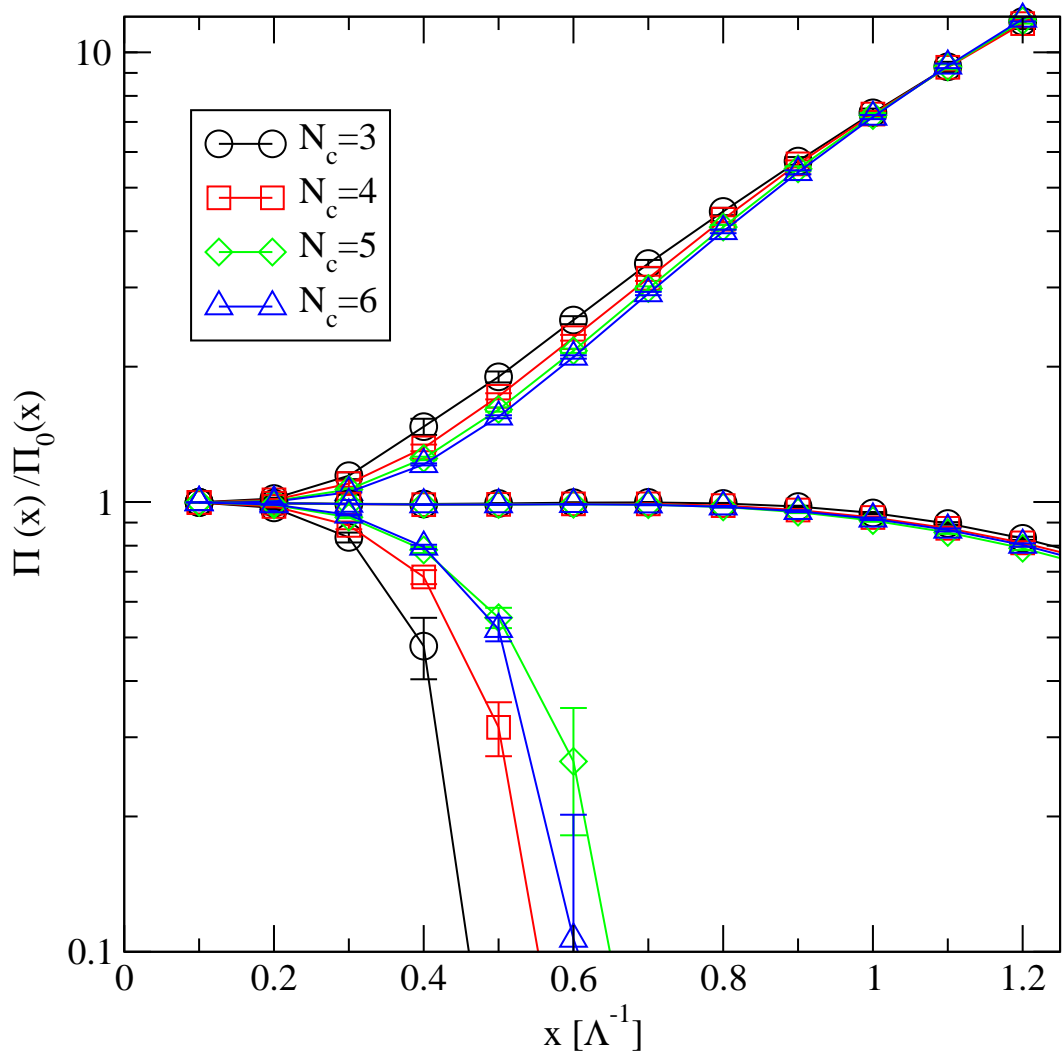} 
\caption{\label{fig_cor}
Meson correlation function in a pure gauge instanton 
ensemble for different numbers of colors. We show the 
correlation function of the pion, the rho meson, and
the $\eta'$ meson normalized to the corresponding 
free correlation functions.}
\end{figure}

\begin{figure}
\centering
\vspace*{3cm}
\includegraphics[width=14cm]{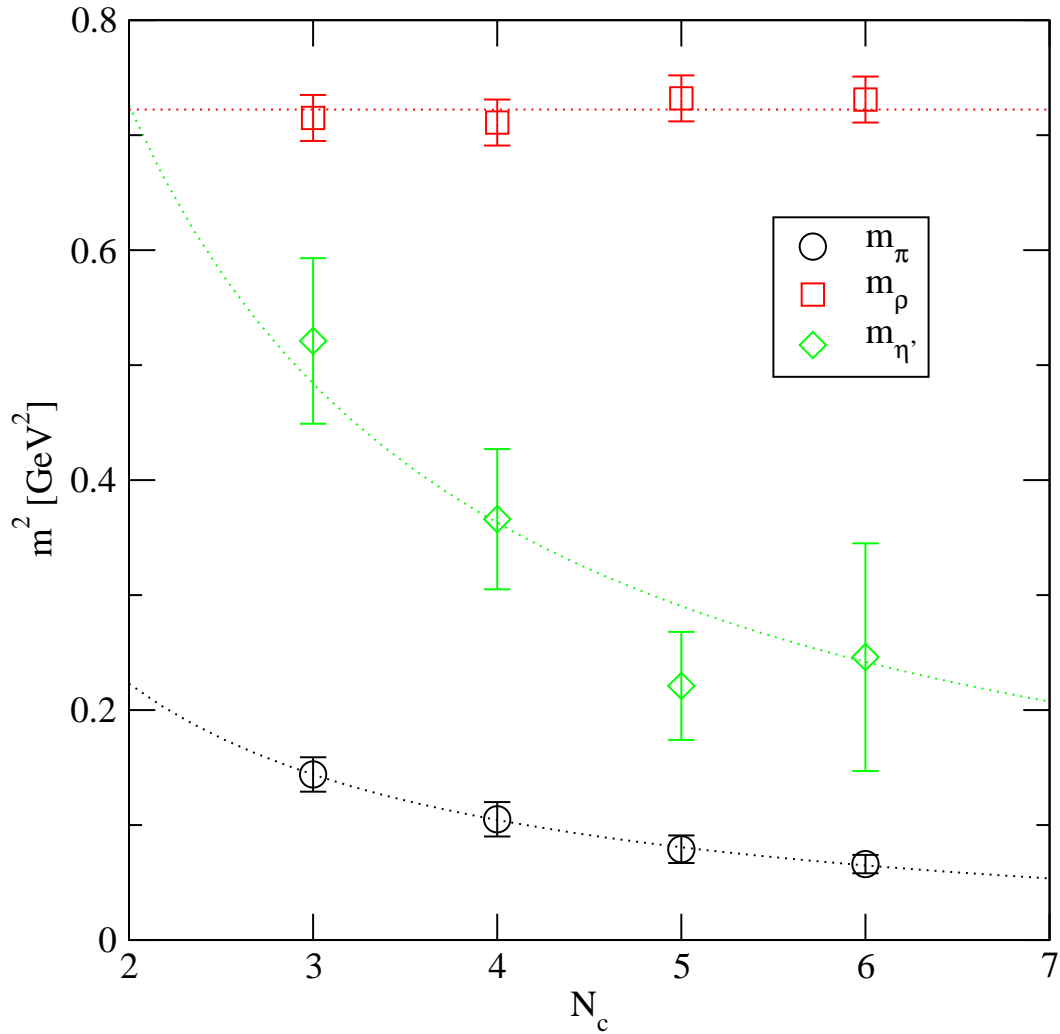} 
\caption{\label{fig_mass}
Masses of the pion, the rho meson, and the $\eta'$ meson 
extracted from meson correlation function in a pure gauge 
instanton ensemble for different numbers of colors. The 
results were converted to physical units using $\Lambda=
0.2$ GeV. The quark mass was chosen to be $m_q=0.2\Lambda
=40$ MeV. The dashed lines show fits of the form $a_1+a_2/N_c$ 
(for $m^2_\pi$ and $m^2_\rho$) and $a_2/N_c$ (for $m^2_{\eta'}$).}
\end{figure}

 The correlation functions measured in numerical 
simulations of the instanton liquid for $N_c=3,
\ldots 6$ are shown in Fig.~\ref{fig_cor}. The meson 
masses extracted from the spectral representation 
equs.~(\ref{spec_pi},\ref{spec_rho},\ref{spec_eta})
are shown in Fig.~\ref{fig_mass}. The results were
obtained from simulations with $N=128$ instantons 
in a euclidean volume $V\Lambda^4=V_3\times 5.76$.
$V_3$ was adjusted such that $(N/V)=(N_c/3)\Lambda^4$.
In order to avoid finite volume artifacts the current
quark mass was taken to be rather large, $m_q=0.2
\Lambda$. We observe that the rho meson correlation 
function exhibits almost perfect scaling
with $N_c$ and as a result the rho meson mass is 
practically independent of $N_c$. The scaling is 
not as good in the case of the pion. As a consequence
there is some variation in the pion mass. However,
as one can see from the fit shown in Fig.~\ref{fig_mass},
this effect is consistent with $1/N_c$ corrections
that amount to about 40\% of the pion mass for
$N_c=3$. Finally, we study the behavior of the $\eta'$
correlation function. There is a clear tendency 
toward $U(1)_A$ restoration, but the correlation
function is still very repulsive for $N_c=6$. As one
can see from Fig.~\ref{fig_mass} the result is consistent
with $m_{\eta'}^2\sim 1/N_c$ although the error bars 
are quite large. 

\begin{figure}
\centering
\vspace*{3cm}
\includegraphics[width=14cm]{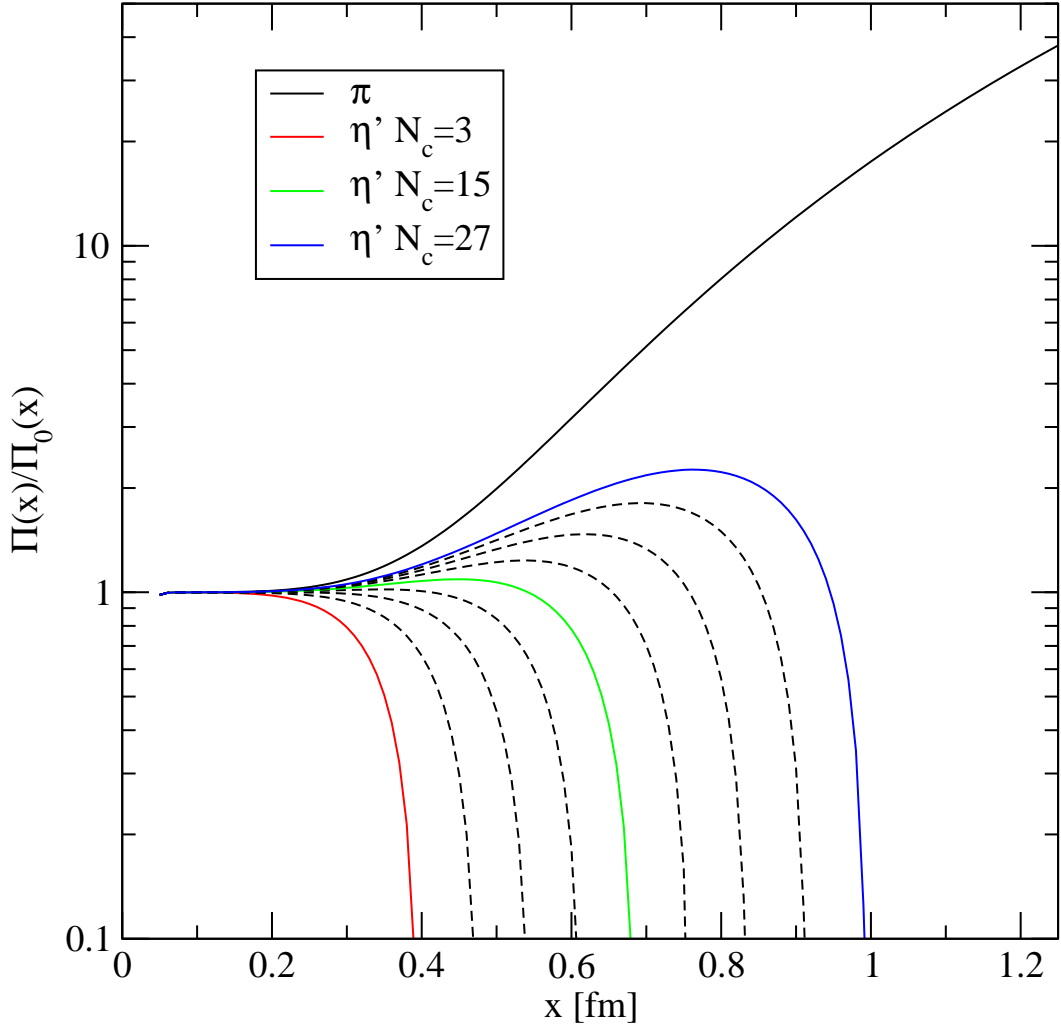} 
\caption{\label{fig_eta}
Expected behavior of the quenched pion and $\eta'$ correlation
functions in the large $N_c$ limit. The correlation functions
were computed from the spectral representation 
equs.~(\ref{spec_pi},\ref{spec_eta}) using the phenomenological
values $m_\pi=139$ MeV, $\lambda_\pi=(450\,{\rm MeV})^2$, 
$E_0=1.3$ GeV and $m_0=900$ MeV ($N_c=3$). We assume that 
$\lambda_\pi^2\sim N_c$ and $m_0^2\sim 1/N_c$.}
\end{figure}

 For comparison, we show the expected behavior of the 
correlation functions based on standard large $N_c$ counting 
in Fig.~\ref{fig_eta}. We have used the spectral representation
equs.~(\ref{spec_pi},\ref{spec_eta}) together with the 
phenomenological values $m_\pi=139$ MeV, $\lambda_\pi=
(450\,{\rm MeV})^2$, $E_0=1.3$ GeV and $m_0=900$ MeV 
($N_c=3$). We assume that $\lambda_\pi^2\sim N_c$ and 
$m_0^2\sim 1/N_c$. We observe that the $\eta'$ correlation
function only approaches the pion correlation for fairly
large values of $N_c$. For example, the $\eta'$ correlation
function does not show intermediate range attraction unless
$N_c>15$. The variation in going from $N_c=3$ to $N_c=6$
is not dramatic, in agreement with the results shown 
in Fig.~\ref{fig_cor}.

\section{Summary}
\label{sec_sum}
 
 In summary we have studied instantons in the large $N_c$
limit of QCD. We have argued that it is possible for the 
instanton liquid model to have a smooth large $N_c$ limit
which is in agreement with scaling relations derived 
from Feynman diagrams. In this limit the density of instantons
grows as $N_c$ whereas the typical instanton size remains
finite. Interactions between instanton are important 
and suppress fluctuations of the topological charge.
As a result the $U(1)_A$ anomaly is effectively restored
even though the number of instantons increases. Using mean 
field arguments \cite{Diakonov:1983hh} and numerical simulations 
we have shown that this scenario does not require fine tuning. 
It arises naturally if the instanton ensemble is stabilized by 
a classical repulsive core. In this case we obtain a picture
in which the instanton density is large but the instanton
liquid remains dilute because instantons are not strongly
overlapping in color space. Further investigations will
have to show whether this scenario is indeed correct. 
For example, it would be useful to study the exact moduli
space for multi-instanton configurations in the large 
$N_c$ limit.

 Of course, there is no a priori reason why instantons
have to be compatible with standard large $N_c$ counting
rules. Instantons are not part of the diagrammatic expansion
and do not need to satisfy scaling relations derived from
diagrams. On the other hand it would be hard to reconcile
the quantitative success of the instanton liquid model in 
describing chiral symmetry breaking and the mass of the 
$\eta'$ with the phenomenological success of the $1/N_c$
expansion if instanton effects strongly violate $N_c$ 
counting rules. 

 We should emphasize that the numerical results presented
in this work only cover fairly small values of $N_c$, $N_c\leq 6$,
and that all the analytical results were obtained in the mean 
field approximation. We cannot exclude the possibility 
that there is a phase transition as the number of colors becomes 
large \cite{Neuberger:1980as,Gross:1994mr,Carter:2001ih}. Carter
and Shuryak suggested, for example, that for clusters involving
$O(N_c)$ instantons $1/N_c$ suppressed color singlet exchanges
become dominant and lead to the formation of tightly bound 
molecules. We did not observe this phenomenon in our simulations
even if a short range color singlet interaction was included, 
but the number of colors ($N_c\leq 10$) may have been too small.
We also did not investigate the possibility that instantons in 
the large $N_c$ limit melt or dissociate into constituents with 
fractional topological charge. The latter scenario was investigated 
in the case of the $CP^{N-1}$ model in \cite{Diakonov:1999ae}.

\begin{figure}
\centering
\includegraphics[width=14cm]{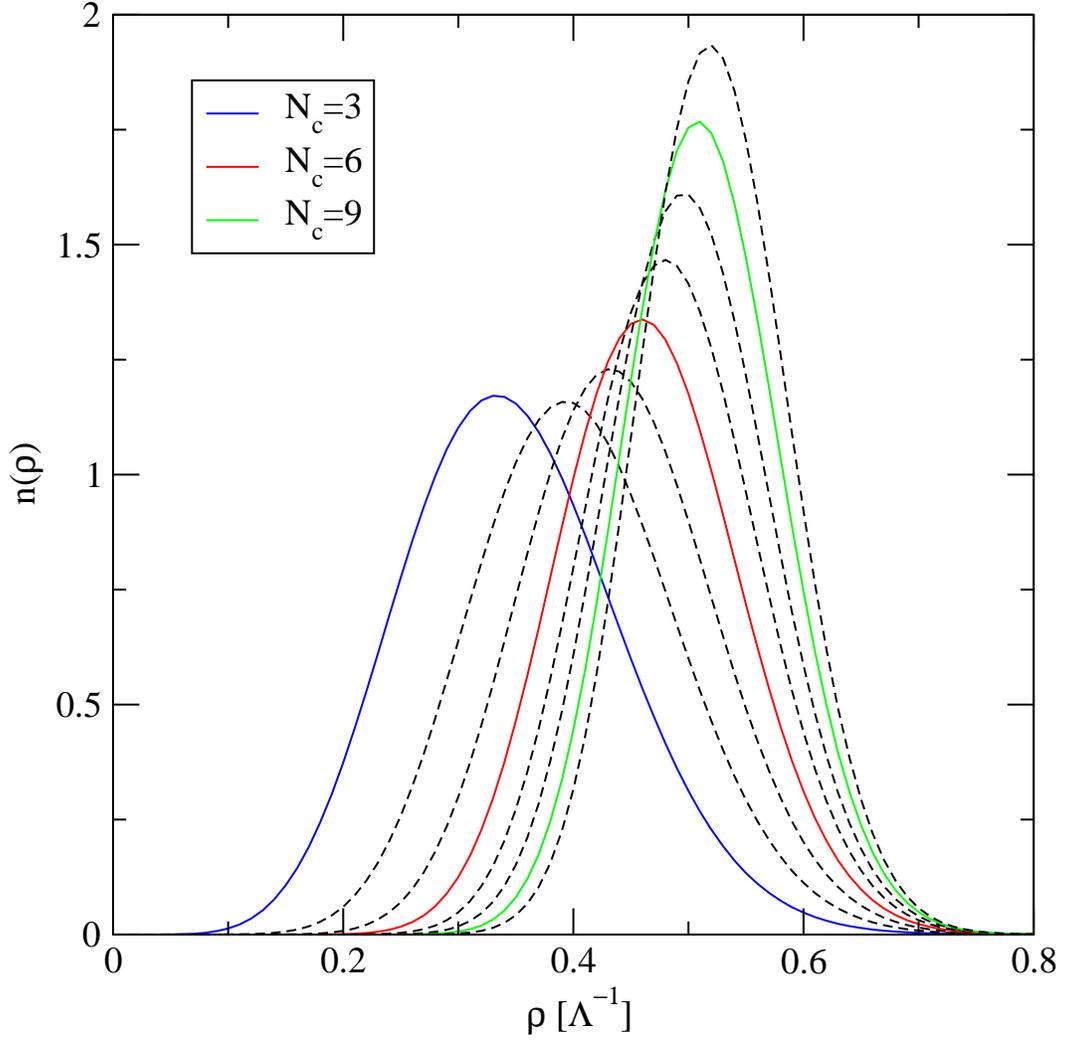}  
\caption{\label{fig_mfa_rho}
Instanton size distribution $n(\rho)$ for different numbers
of colors $N_c=3,\ldots,10$. The results show in this figure
were obtained using the mean field approximation. }
\end{figure}

\begin{figure}
\centering
\vspace*{3cm}
\includegraphics[width=14cm]{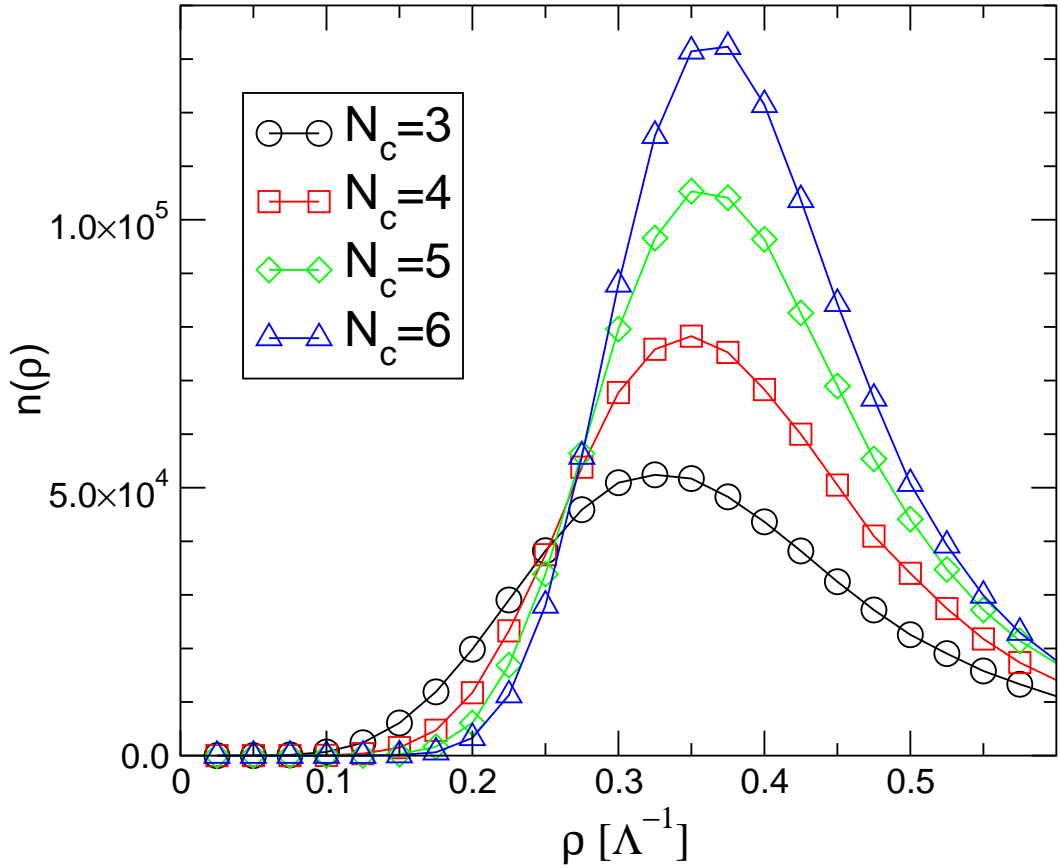} 
\caption{\label{fig_nrho}
Instanton size distribution in a pure gauge instanton 
ensemble for different numbers of colors. The results
were obtained using numerical simulations with $N=128$
instantons. }
\end{figure}

\begin{figure}
\centering
\vspace*{3cm}
\includegraphics[width=14cm]{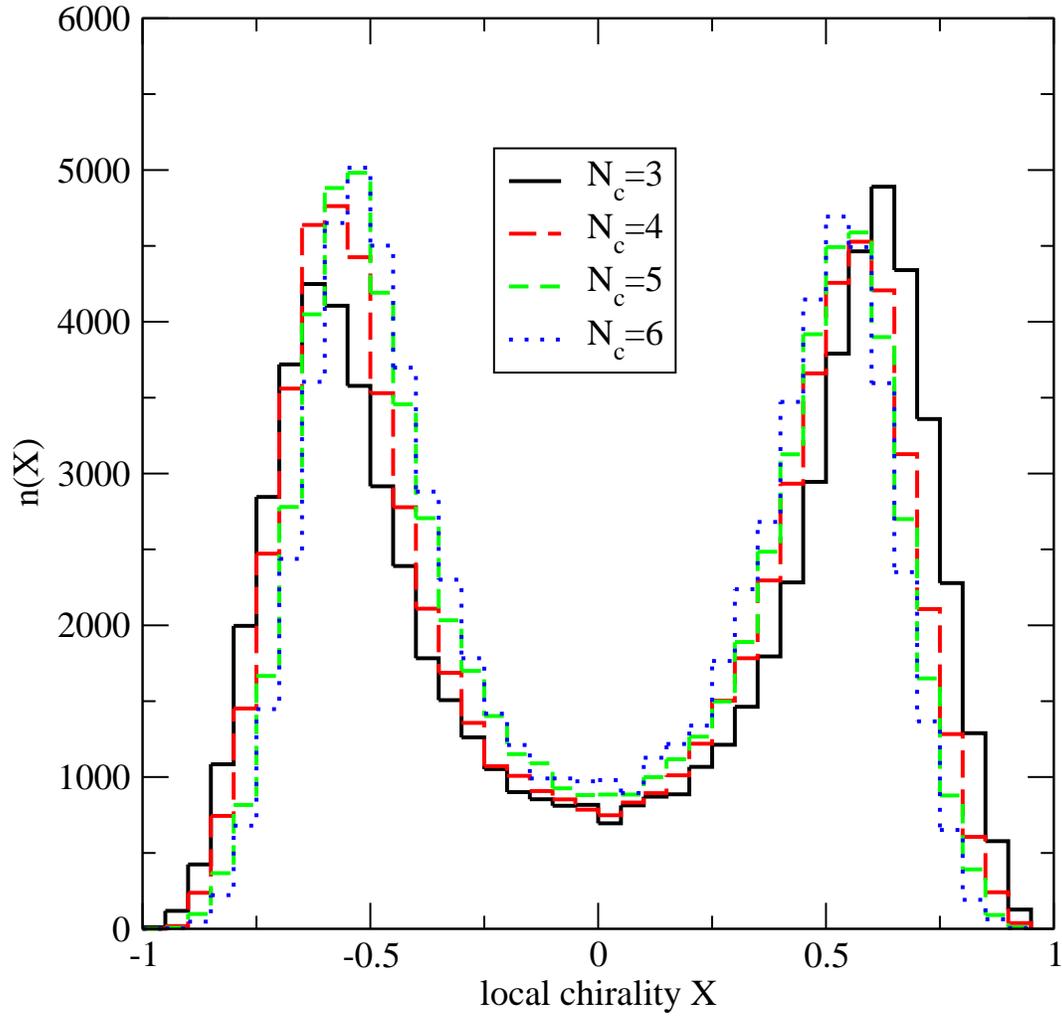} 
\caption{\label{fig_isgur}
Local chirality distribution function in a pure gauge instanton 
ensemble for different numbers of colors. The local chirality 
$X$ is defined in the text. The results were obtained using 
numerical simulations with $N=128$ instantons. }
\end{figure}

 Our results can be compared to the lattice results 
of Lucini and Teper \cite{Lucini:2001ej} and
Cundy, Teper and Wenger \cite{Cundy:2002hv}. Lucini
and Teper find a fixed point in the instanton 
size distribution, in agreement with our results
shown in Figs.~\ref{fig_mfa_rho} and \ref{fig_nrho}.
However, they do not find any suppression of large
size instantons. Cundy et al.~studied the local
chirality distribution. They find that the double
peak structure of this distribution disappears in
the large $N_c$ limit. Our results shown in 
Fig.~\ref{fig_isgur} are in qualitative, but 
not in quantitative agreement with their findings. 
In our calculations the effect is much less 
pronounced. This implies that either the instanton 
liquid is not as dilute as it is in our calculations, 
that mixing with non-zero modes becomes more important 
as $N_c$ increases, or that instantons are no longer 
semi-classical.

Acknowledgments: I would like to thank D.~Diakonov, D.~Gross, 
E.~Shuryak, U.~Wenger and A.~Zhitnitsky for useful discussions. 
I would also like to acknowledge the hospitality of the Institute
for theoretical Physics at UCSB where this work was completed.
This work was supported in part by US DOE grant DE-FG-88ER40388
and by the National Science Foundations under grant 
No.~PHY99-07949.

\newpage

\newpage

\newpage

\newpage 

\newpage 

\newpage 

\newpage 

\newpage 

\newpage 

\newpage 

\newpage 

\newpage 

\newpage 

\newpage 

\newpage 

\newpage 

\end{document}